\documentclass[preprint,3p,sort&compress]{elsarticle}
\usepackage{graphicx}
\usepackage{epsfig}
\usepackage[intlimits,centertags]{amsmath}
\usepackage{amssymb,amsfonts}
\usepackage{gensymb}
\usepackage{mathrsfs}

\medskipamount
\newcommand{\D}{\mathrm{d}}

\begin{document}
\vspace*{-0.3cm}
\hspace*{13cm} \hbox{\rm YITP-SB-10-014}
\vskip 1cm
\title{GZK Neutrinos after the Fermi-LAT Diffuse Photon Flux Measurement}
\author[sb]{M.~Ahlers} 
\ead{ahlers@insti.physics.sunysb.edu}
\author[uwm]{L.~A.~Anchordoqui}
\ead{doqui@gravity.phys.uwm.edu}
\author[sb,ub]{M.~C.~Gonzalez--Garcia}
\ead{concha@insti.physics.sunysb.edu}
\author[uw]{F.~Halzen}
\ead{halzen@icecube.wisc.edu}
\author[uo]{S.~Sarkar}
\ead{s.sarkar@physics.ox.ac.uk}
\address[sb]{%
  C.N.~Yang Institute for Theoretical Physics,
  SUNY at Stony Brook, Stony Brook, NY 11794-3840, USA}
\address[uwm]
{Department of Physics, University of
  Wisconsin-Milwaukee, Milwaukee, WI 53201, USA} 
\address[ub]{
  Instituci\'o Catalana de Recerca i Estudis Avan\c{c}ats (ICREA),
  Departament d'Estructura i Constituents de la Mat\`eria and ICC-UB, 
  Universitat
  de Barcelona, 647 Diagonal, E-08028 Barcelona, Spain}
\address[uw]{%
Department of Physics, University of Wisconsin, Madison, WI 53706,USA}
\address[uo]{Rudolf Peierls Centre for Theoretical Physics, University
  of Oxford, Oxford OX1 3NP, UK}

\begin{abstract}
  Cosmogenic neutrinos originate from photo-hadronic interactions of
  cosmic ray protons with the cosmic microwave background (CMB).  The
  neutrino production rate can be constrained through the accompanying
  electrons, positrons and $\gamma$-rays that quickly cascade on the CMB
  and intergalactic magnetic fields to lower energies and generate a
  $\gamma$-ray background in the GeV-TeV region.  Bethe-Heitler pair
  production by protons also contributes to the cascade and can
  tighten the neutrino constraints in models where extragalactic
  cosmic rays begin to dominate over the galactic component at a
  relatively low ``crossover'' energy.  We investigate this issue in
  the light of the recent Fermi-LAT measurements of the diffuse
  extragalactic $\gamma$-ray background and illustrate by a fit to the
  HiRes spectrum how the prediction of the cosmogenic neutrino flux in
  all-proton models varies with the crossover energy.  The neutrino
  flux is required to be smaller when the $\gamma$-ray bound is applied,
  nevertheless such models are still consistent with HiRes and
  Fermi-LAT if one properly takes into account the energy uncertainty
  of cosmic ray measurements. The presently allowed flux is within
  reach of the IceCube neutrino telescope and other dedicated radio
  experiments.
\end{abstract}

\maketitle

\newpage
\section{Introduction}

Soon after the discovery of the cosmic microwave background
(CMB)~\cite{Penzias:1965wn}, it was realized that interactions of
extragalactic ultrahigh energy (UHE) cosmic rays (CRs) on the relic
photons would suppress the cosmic ray flux at energies $\gtrsim 5
\times 10^{10}$~GeV, the so-called ``GZK
cutoff''~\cite{Greisen:1966jv,Zatsepin:1966jv}. It was pointed out
subsequently~\cite{Beresinsky:1969qj} that the GZK interaction also
generates a ``cosmogenic flux'' of neutrinos, through the decay of
secondary charged pions. Forty years later, the predicted suppression
of the UHE CR flux was indeed observed by the
HiRes~\cite{Abbasi:2007sv} and Auger~\cite{Abraham:2008ru}
experiments. However, the cosmogenic flux of neutrinos has yet to be
detected.

The GZK reaction chain generating cosmogenic neutrinos is well
known~\cite{Stecker:1978ah}. The intermediate state of the reaction $p
\gamma_{\rm CMB} \to n \pi^+/p\pi^0$ is dominated by the $\Delta^+$
resonance, because the neutron decay length is smaller than the
nucleon mean free path on the CMB. Resonant $p\gamma$ interactions
produce twice as many neutral pions as charged pions.  Direct pion
production via virtual meson exchange contributes only about 20\% to
the total cross-section, but is almost exclusively into
$\pi^+$. Hence, $p\gamma$ interactions produce roughly equal number of
$\pi^+$ and $\pi^0$. Gamma-rays, produced via $\pi^0$ decay,
subsequently cascade electromagnetically on intergalactic radiation
fields through $e^+ e^-$ pair production followed by inverse Compton
scattering. The net result is a pile up of $\gamma$-rays at GeV-TeV
energies, just below the threshold for further pair production on the
diffuse optical background. Meanwhile each $\pi^+$ decays to 3
neutrinos and a positron; the $e^+$ readily loses its energy through
inverse Compton scattering on the diffuse radio background or through
synchrotron radiation in intergalactic magnetic fields. The neutrinos
carry away about 3/4 of the $\pi^+$ energy, therefore the energy in
cosmogenic neutrinos is about 3/4 of that produced in $\gamma$-rays.

The normalization of the neutrino flux depends critically on the
cosmological evolution of the CR sources and on their proton injection
spectra~\cite{Yoshida:pt}. It also depends on the assumed spatial
distribution of sources; for example, local sources in the Virgo
cluster~\cite{Hill:1985mk}, would dominate the high energy tail of the
proton spectrum.  Another source of uncertainty is the energy at which
there is a transition from Galactic to extragalactic CRs as inferred
from a change in the spectral slope.  The ``ankle'' at $\sim 3
\times10^9$~GeV seems to be a natural candidate for this
transition~\cite{Linsley:1963bk,Hill:1983mk,Wibig:2004ye}, but a lower
energy crossover at the ``second knee'' at $\sim 5 \times10^8$~GeV has
also been advocated~\cite{Berezinsky:2002nc,Fodor:2003ph}.  A fourth
source of uncertainty is the chemical composition of the parent CRs --
if these are heavy nuclei rather than protons, then the neutrino flux
is reduced~\cite{Hooper:2004jc}.
 
The most up-to-date calculation~\cite{Anchordoqui:2007fi} of the
cosmogenic neutrino flux combines a double-fit analysis of the
energy~\cite{Abraham:2010mj} and elongation rate~\cite{Abraham:2010yv}
measurements to constrain the spectrum and chemical composition of UHE
CRs at their sources. Injection models with a wide range of chemical
compositions are found to be consistent with observations. The data is 
consistent with a proton-dominated spectrum with a small admixture of 
heavy nuclei, in which case the cosmogenic neutrino flux is rather similar 
to the all-proton model. In this case, kilometer-scale neutrino telescopes 
are expected to observe of ${\cal O}(1)$ cosmogenic neutrino event per year. 
In contrast, an intermediate to heavy nuclear composition beyond the ankle, 
as indicated by the elongation rate vs.~energy~\cite{Abraham:2010yv}, can
lead to a considerable suppression (up to two orders of magnitude) of
the cosmogenic neutrino flux in comparison to the all-proton case. 
However, little is known about the chemical composition from just below to 
beyond the GZK cutoff, where the most significant contribution to 
cosmogenic neutrinos form  UHE CR protons is expected. 
It is interesting to note that uncertainties in the
extrapolation of the proton-air interaction -- cross-section,
elasticity and multiplicity of secondaries -- from accelerator
measurements to the high energies characteristic for air showers
are large enough to undermine any definite conclusion on the chemical
composition~\cite{Wibig:2008ji,Ulrich:2009yq}.

In this work we study the constraint set by the diffuse $\gamma$-ray
background on all-proton models of extragalactic CRs.  We parametrize
our ignorance of the crossover energy -- which marks the transition
between the galactic and extragalactic components -- as a variable
low energy cutoff in the proton injection rate. By fitting only to CR
data above the crossover energy, taken to be between $10^{17.5}$~eV
and $10^{19}$~eV, we determine the statistically preferred values of
the spectral index $\gamma$ and cosmic source density evolution index
$n$ by a goodness-of-fit (GOF) test of the HiRes data, taking into
account the energy resolution of about 25\%. For each model we check
that the total energy density of the EM cascade is below a critical
value inferred from the recent measurement of the extragalactic
$\gamma$-ray background by the Fermi-LAT
Collaboration~\cite{Abdo:2010nz}.  We find that the allowed range of
the cosmogenic neutrino flux increases with the crossover energy and
can be up to an order of magnitude {\em larger} than the values
presented in a recent study~\cite{Berezinsky:2010xa}.

This paper is organized as follows. We begin in \S~\ref{sec:I} with a
discussion of the extragalactic proton fluxe and the corresponding
energy density of cascade $\gamma$-rays. We present our statistical
method in \S~\ref{sec:II} and discuss our results in \S~\ref{sec:III}

\section{Extra-galactic Proton Fluxes and Diffuse Gamma Background}
\label{sec:I}

For a spatially homogeneous distribution of cosmic sources, emitting
UHE particles of type $i$, the co-moving number density $Y_i$ is
governed by a set of (Boltzmann) continuity equations of the form:
\begin{equation}\label{diff0}
\dot Y_i = \partial_E(HEY_i) + \partial_E(b_iY_i)-\Gamma_{i}\,Y_i
+\sum_j\int{\rm d} E_j\,\gamma_{ji}Y_j+\mathcal{L}_i\,,
\end{equation}
together with the Friedman-Lema\^{\i}tre equations describing the
cosmic expansion rate $H(z)$ as a function of the redshift
$z$.\footnote{This is given by \mbox{$H^2 (z) = H^2_0\,[\Omega_{\rm
      m}(1 + z)^3 + \Omega_{\Lambda}]$}, normalised to its value today
  of $H_0 \sim70$ km\,s$^{-1}$\,Mpc$^{-1}$, in the usual ``concordance
  model'' dominated by a cosmological constant with $\Omega_{\Lambda}
  \sim 0.7$ and a (cold) matter component, $\Omega_{\rm m} \sim
  0.3$~\cite{Amsler:2008zzb}. The time-dependence of the redshift can
  be expressed via ${\rm d}z = -{\rm d} t\,(1+z)H$.}  The first and
second terms on the r.h.s.~describe, respectively, redshift and other
continuous energy losses (CEL) with rate $b \equiv
\mathrm{d}E/\mathrm{d}t$.  The third and fourth terms describe more
general interactions involving particle losses ($i \to$ anything) with
interaction rate $\Gamma_i$, and particle generation of the form $j\to
i$ with differential interaction rate $\gamma_{ij}$.  The last term on
the r.h.s., $\mathcal{L}_i$, corresponds to the emission rate per
co-moving volume of CRs $i$. We refer
to Ref.~\cite{Ahlers:2009rf} for explicit definitions of the
coefficients in Eq.~(\ref{diff0}).

Extragalactic protons lose their energy via Bethe-Heitler (BH) pair
production and photo-hadronic interactions on cosmic radiation
backgrounds, notably the CMB. Bethe-Heitler pair production,
$p+\gamma_\mathrm{bgr} \to p + e^+ + e^-$, can be treated as a
continuous energy loss due to its low
inelasticity~\cite{Blumenthal:1970nn}. This process dominates the
evolution of the spectra at energies between $10^9$~GeV and a few
times $10^{10}$~GeV. At higher energies, resonant photo-hadronic
interactions with CMB photons lead to a sharp suppression of the
spectrum~\cite{Greisen:1966jv,Zatsepin:1966jv}. The produced charged
and neutral pions release electrons, positrons, neutrinos and photons
through their decay. We calculate the spectra of hadrons and neutrinos
using the Monte Carlo package {\tt SOPHIA}~\cite{Mucke:1999yb}.

It is possible to approximate the energy loss in the hadronic cascade
due to photo-pion production as a CEL with
\begin{equation}\label{eq:CELpgamma}
\frac{{\rm d} E}{{\rm d} t}(z,E) \equiv b(z,E) 
\simeq E\,\Gamma_{p}(z,E) - \int{\rm
d}E'E'\gamma_{pp}(z,E,E')\,.
\end{equation}
Diffractive $p\gamma$ processes at high energies with large final
state multiplicities of neutrons and protons ultimately invalidate the
CEL approximation. However, the relative error below $10^{12}$~GeV is
less than $15$\% so we will use this approximation for a detailed
numerical scan in the model space of proton spectra.

Electromagnetic (EM) interactions of photons and leptons with the
extragalactic radiation backgrounds and magnetic field can happen on
time-scales much shorter than their production rates. The relevant
processes with background photons contributing to the differential
interaction rates $\gamma_{ee}$, $\gamma_{\gamma e}$ and $\gamma_{e
  \gamma}$ are inverse Compton scattering (ICS), $e^\pm+\gamma_{\rm
  bgr}\to e^\pm+\gamma$, pair production (PP), $\gamma+\gamma_{\rm
  bgr}\to e^++e^-$, double pair production (DPP) $\gamma+\gamma_{\rm
  bgr}\to e^++e^-+e^++e^-$, and triple pair production (TPP),
$e^\pm+\gamma_{\rm bgr}\to
e^\pm+e^++e^-$~\cite{Blumenthal:1970nn,Blumenthal:1970gc}. High energy
electrons and positrons can also lose energy via synchrotron radiation
on the intergalactic magnetic field~\footnote{Since we consider a
  relatively strong intergalactic magnetic field, we can neglect TPP
  by electrons in the following~\cite{Lee:1996fp}. Also below
  $10^{12}$~GeV we can safely neglect DPP of photons in the
  calculation~\cite{Demidov:2008az}.} the strength of which is limited
to be below $\sim 10^{-9}$G~\cite{Kronberg:1993vk} and suggested to be
of ${\cal O}(10^{-12})$G by simulations of large-scale structure
formation~\cite{Dolag:2004kp}; details on
the calculation are given in~\ref{app:I}.

Synchrotron radiation in strong magnetic fields can also by-pass the 
EM cascade and transfer energy to sub-GeV photons that are unconstrained 
by the Fermi-LAT spectrum~\cite{Wdowczyk:1972}. In the case of a strong
$10^{-9}$G field this can be relevant for electrons around $10^{9}$~GeV, 
where synchrotron loss starts to dominate over ICS loss in the CMB and 
where the corresponding synchrotron spectrum still peaks below 100~MeV.
In our calculation we adopt a moderate value of $10^{-12}$G following~\cite{Dolag:2004kp}. 
However, we have checked that a significantly larger field strength of 
$10^{-9}$G has little effect on the $\gamma$-ray flux in the Fermi-LAT 
energy range relevant for our discussion (see Fig.~\ref{fig:magplot} 
in the Appendix).

In this paper the emission rate of CR protons per co-moving volume
(GeV${}^{-1}$ cm${}^{-3}$ s${}^{-1}$) is assumed, as per usual
practice, to follow a power-law:
\begin{equation}
\mathcal{L}_p(0,E) \propto (E/E_0)^{-\gamma}\times\begin{cases} f_
-(E/E_{\rm min})&E<E_{\rm min}\,,\\1& E_{\rm min}<E<E_{\rm max}\,,\\
f_+(E/E_{\rm max})& E_{\rm max}<E\,.
\end{cases}
\label{eq:injection}
\end{equation}
We will consider spectral indices $\gamma$ in the range $2\div3$. The
functions $f_\pm(x) \equiv x^{\pm2}\exp(1-x^{\pm2})$ in
Eq.~(\ref{eq:injection}) smoothly turn off the contribution below
$E_{\rm min}$ and above $E_{\rm max}$. We set $E_{\rm
  max}=10^{21}$~eV in the following and vary $E_{\rm min}$ in the
range $10^{17.5}\div10^{19}$~eV, corresponding to a
galactic-extragalactic crossover between the ``second knee'' and the
``ankle'' in the CR spectrum.

The cosmic evolution of the spectral emission rate per comoving volume
is parameterized as:
\begin{equation}
\mathcal{L}_p(z,E) = \mathcal{H}(z)\mathcal{L}_p(0,E)\,.
\label{eq:sourden1}
\end{equation}
For simplicity, we use the standard approximation
\begin{equation}
\mathcal{H}(z) \equiv (1+z)^n\Theta(z_{\rm max}-z)\,,
\label{eq:sourden2}
\end{equation} 
with $z_{\rm max} = 2$. Note that the dilution of the source density
due to the Hubble expansion is taken care of since $\mathcal{L}$ is
the {\em comoving} density, {\it i.e.} for no evolution we would
simply have $\mathcal{H} = 1$. We consider cosmic evolution of UHE CR
sources with $n$ in the range $2\div6$.


As mentioned above, EM interactions of photons and
leptons with the extra-galactic background light and magnetic field
can happen on time-scales much shorter than their production rates.
It is convenient to account for these contributions during the proton
propagation as fast developing electro-magnetic cascades at a fixed
redshift. We will use the efficient method of ``matrix
doubling''~\cite{Protheroe:1992dx} for the calculation of the
cascades. Since the cascade $\gamma$-ray flux is mainly in the GeV-TeV
region and has an almost universal shape here, it is numerically much
more efficient to calculate the total energy density $\omega_{\rm
  cas}$ injected into the cascade and compare this value to the limit
imposed by Fermi-LAT.  The total energy density (eV cm${}^{-3}$) of
EM radiation from proton propagation in the past is given
as
\begin{equation}
\label{eq:omegacas}
\omega_{\rm cas} \equiv \int{\rm d}E E n_{\rm cas}(0,E) = \int{\rm
d}t\int{\rm d}E \,\frac{b_{\rm cas}(z,E)}{(1+z)^4}\,n_{\rm p}(z,E)\,,
\end{equation}
where $n(z,E)$ is the physical energy density at redshift $z$, defined
via $n(z,E)\equiv(1+z)^3Y(z,E)$.  We discuss the derivation of this
equation in~\ref{app:II}. The continuous energy loss of protons into
the cascade, denoted by $b_{\rm cas}$, is in the form of electron,
positron and $\gamma$-ray production in BH ($b_{\rm BH}$) and
photo-pion ($b_\pi$) interactions.\footnote{Note the difference
  between our Eq.~(\ref{eq:omegacas}) and Eq.~(10) in
  Ref.~\cite{Berezinsky:2010xa} where the approximation
  $\partial_Eb(z,E) \simeq b(z,E)/E$ is used ({\it cf.}~Fig.~5 in
  Ref.~\cite{Ahlers:2009rf}) and an adiabatic scaling with redshift,
  $b(z,E) \simeq (1+z)^2b(0,E(1+z))$, is assumed.}.

In the following we derive the BH and photo-pion contribution to
$\omega_{\rm cas}$ separately. For the photo-pion contribution we
estimate
\begin{equation}\label{eq:bpion}
b_{\pi}(z,E) \simeq \int{\rm d}E' E'\left[\gamma_{pe^-}(z,E,E') +
\gamma_{pe^+}(z,E,E') + \gamma_{p\gamma}(z,E,E')\right]
\end{equation}
For the energy loss via BH pair production we use the expression given
in Ref.~\cite{Blumenthal:1970nn}.  Note, that since the photo-pion
contribution in the cascade is dominated at the GZK cutoff its
contribution should {\it increase} relative to BH pair production with
increasing crossover energy and, hence, also the associated neutrino
fluxes after normalization to $\gamma$-ray and CR data.

\section{Goodness of Fit Test}\label{sec:II}

\begin{figure}[t]
\begin{center}
\includegraphics[width=0.49\linewidth]{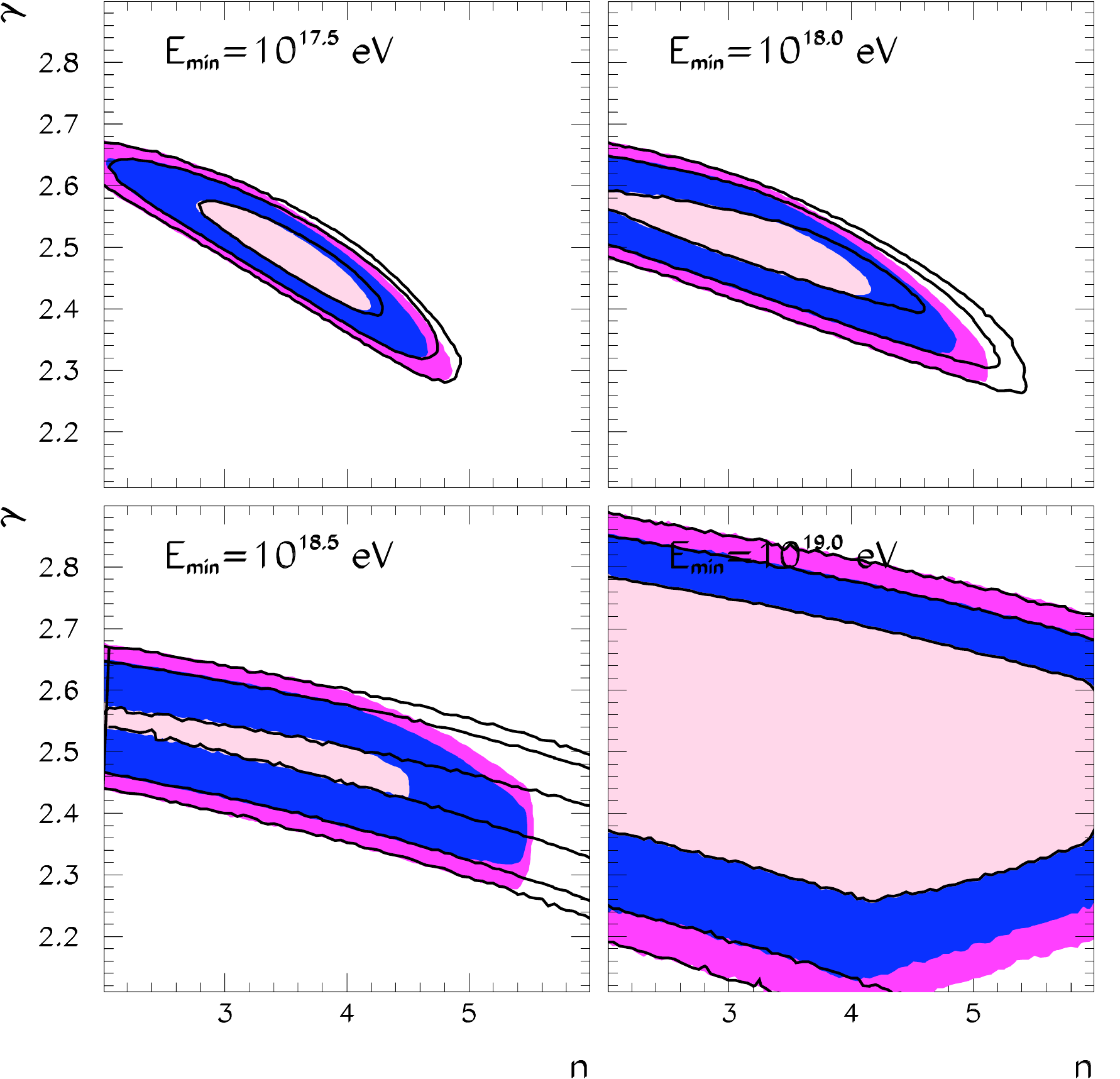}\hfill
\includegraphics[width=0.49\linewidth]{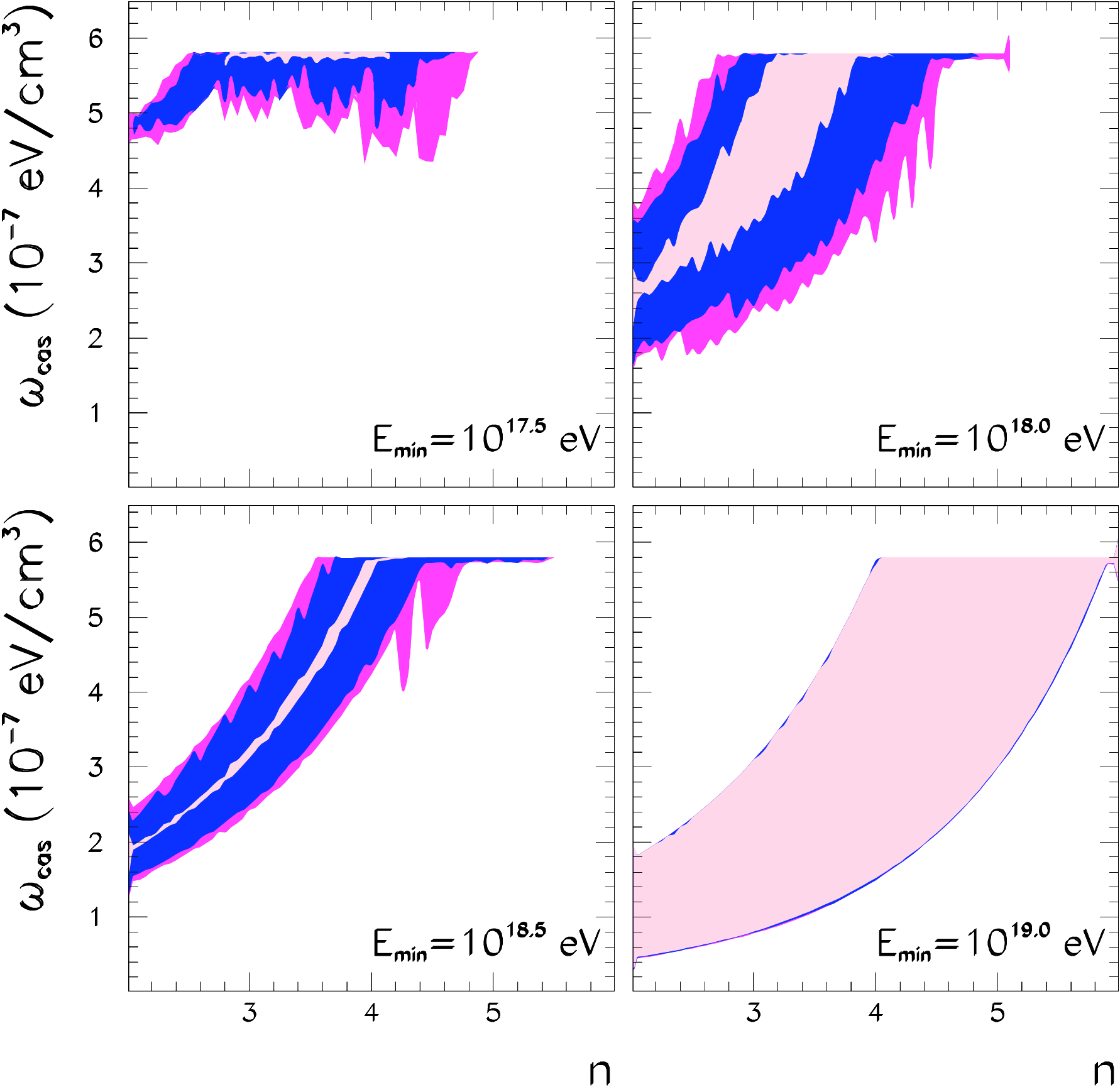}
\end{center}
\vspace{-0.3cm}
\caption[]{{\bf Left Panel:} Goodness of fit test of the HiRes data
  \cite{Abbasi:2007sv}. We show the 68\% (pink), 95\% (blue) and 99\%
  (magenta) confidence levels of the injection index $\gamma$ and the
  cosmic evolution index $n$. The black lines indicate the allowed
  regions before the cascade ($\omega_{\rm cas}$) bound is imposed.
  {\bf Right Panel:} The corresponding energy density in the EM
  cascade.}
\label{fig:gof}
\end{figure}
\begin{figure}[t]
\begin{center}
\includegraphics[width=0.7\linewidth]{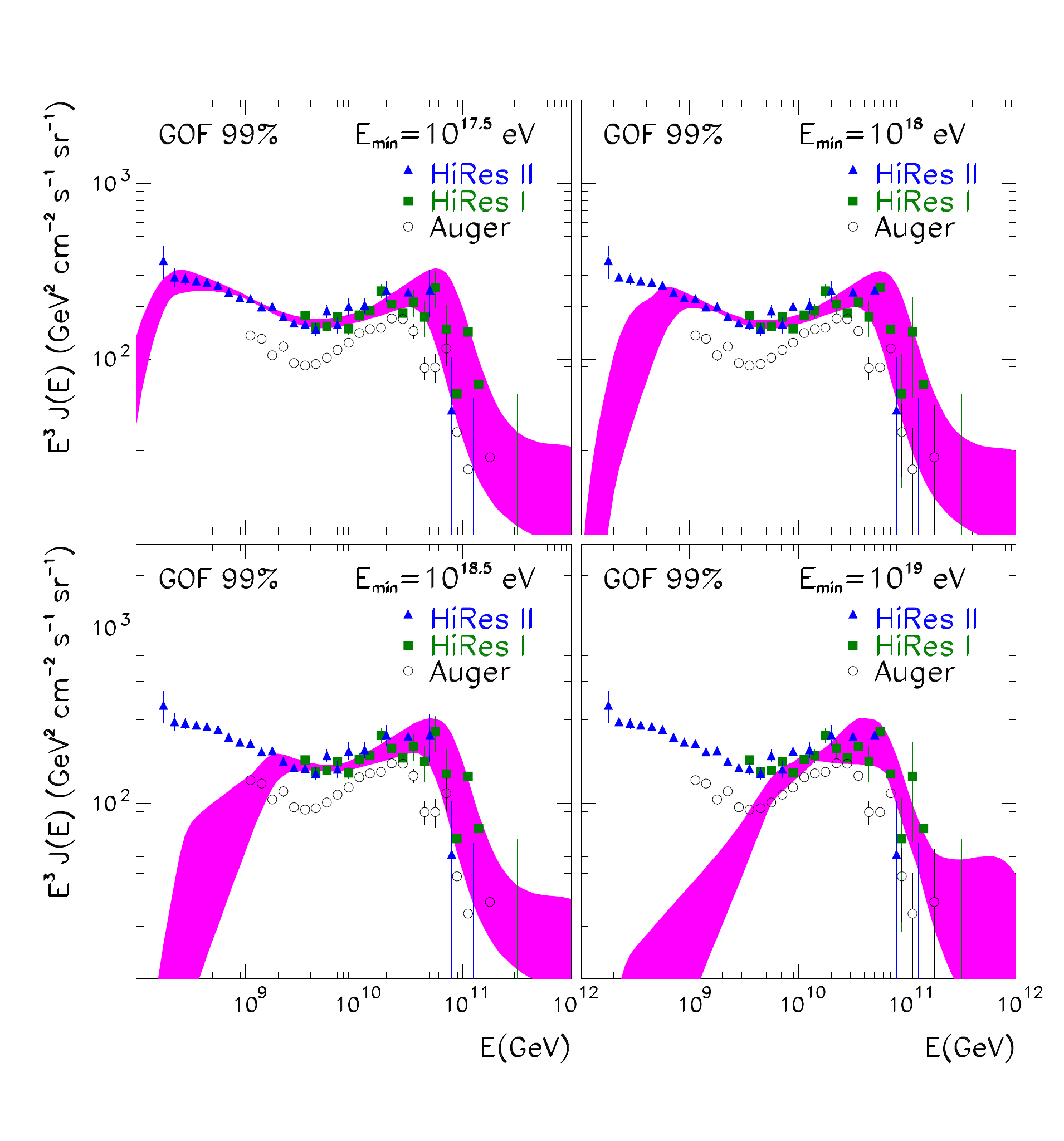}
\end{center}
\vspace{-0.3cm}
\caption[]{The allowed proton flux (at the 99\% confidence level) for
  increasing crossover energy $E_{\rm min}$. Each fit of the proton
  spectrum is marginalized with respect to the experimental energy
  uncertainty and we show the shifted predictions in comparison to the
  HiRes central values~\cite{Abbasi:2007sv}.  For comparison we also
  show the Auger data~\cite{Abraham:2008ru,Abraham:2010mj} which has
  {\em not} been included in the fit.}
\label{fig:omega}
\end{figure}
\begin{figure}[t]
\begin{center}
\includegraphics[width=0.5\linewidth]{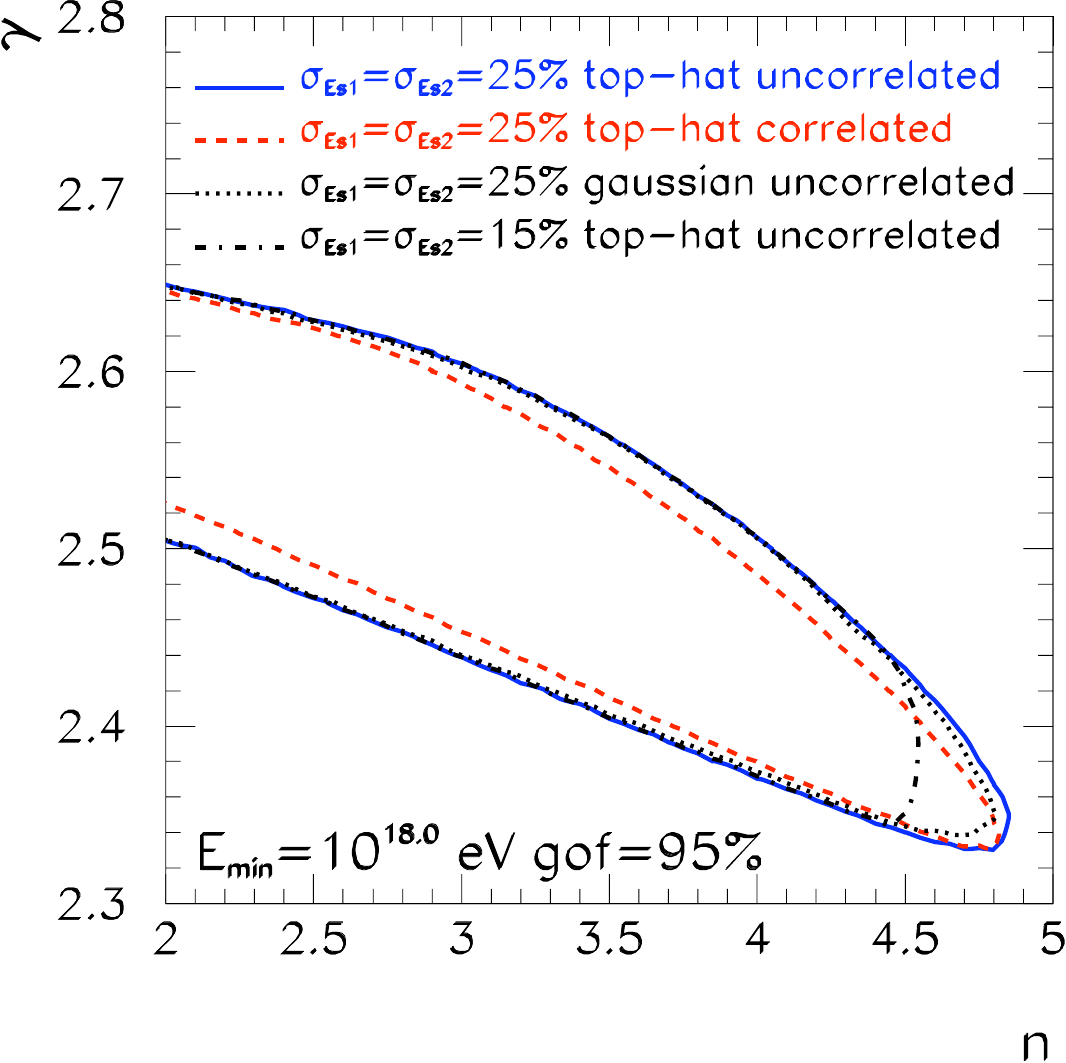}
\end{center}
\vspace{-0.3cm}
\caption[]{Systematic effect of the experimental energy resolution on
  the fitted spectral index $\gamma$ and cosmological evolution
  parameter $n$. For illustration we show the dependence of the 95\%
  C.L. bound for a crossover energy of $10^{18}$~eV. The blue contour
  corresponds to the region shown in Fig.~\ref{fig:gof} assuming an
  uncorrelated energy shift of 25\% in both data sets (HiRes I and
  II)~\cite{Abbasi:2007sv}, for a flat prior (``top-hat''
  distribution). The red dashed curve assumes correlated errors of the
  energy resolution in both data sets. The black dotted curve shows
  the result for uncorrelated errors with a Gaussian prior, and the
  dashed-dotted line shows uncorrelated errors with a flat prior, but
  with a lower uncertainty of 15\%.}
\label{fig:compa}
\end{figure}

In this section we present the results of a GOF test of the
compatibility of a given model, characterized by the injection
spectral index $\gamma$, cosmic evolution index $n$, and crossover
energy $E_{\rm min}$, with the CR experimental data -- in particular
HiRes I and II \cite{Abbasi:2007sv} -- imposing also consistency with
the Fermi-LAT measurements of the diffuse extra-galactic $\gamma$-ray
background.

Given the acceptance $A_i$ (in units of area per unit time per unit
solid angle) of the experiment for the energy bin $i$ centered at
$E_i$ and with bin width $\Delta_i$, and the energy scale uncertainty
of the experiment, $\sigma_{E_s}$ the number of expected events in the
bin is given by
\begin{equation}
N_i(n,\gamma,{\cal N},\delta)=
A_i \int_{E_i(1+\delta)-\Delta_i/2}^{E_i(1+\delta)+\Delta_i/2} 
J^p_{{\cal N},n,\gamma}(E) 
dE\;, 
\label{eq:nev}
\end{equation}
where $J^p_{{\cal N},n,\gamma}(E)=n_p(0,E) \frac{c}{4\pi}$ is the
proton flux arriving at the detector corresponding to a proton source
luminosity as in Eq.~\eqref{eq:injection}, with the cosmic evolution
of the source density given by Eqs.~\eqref{eq:sourden1}
and~\eqref{eq:sourden2}.  The parameter $\delta$ in Eq.~\eqref{eq:nev}
above is a fractional energy-scale shift that reflects the
energy-scale uncertainty of the experiment, and ${\cal N}$ is the
normalization of the proton source luminosity.

The probability distribution of events in the $i$-th bin is of the
Poisson form with mean $N_i$.  Correspondingly the $r$-dimensional
($r$ being the number of bins of the experiment with $E_i\geq E_{\rm
  min}$) probability distribution for a set of non-negative integer
numbers ${\vec k}=\{k_1,...k_r\}$, $P_{\vec k}(n,\gamma,{\cal
  N},\delta)$, is just the product of the individual Poisson
distributions.

According to this $r$-dimensional probability distribution, the
experimental result ${\vec N^{\rm exp}}=\{N^{\rm exp}_1,...,N^{\rm
  exp}_r\}$ has a probability $P_{\vec N^{\rm exp}}(n,\gamma,{\cal
  N},\delta)$ and correspondingly the experimental probability after
marginalizing over the energy scale uncertainty and normalization is:
\begin{equation}
P_{\rm exp}(n,\gamma)={\rm Max}_{\delta,{\cal N}} 
P_{\vec N^{\rm exp}}(n,\gamma,{\cal N},\delta)\,. 
\label{eq:margi}
\end{equation}
where the maximization is made within some prior for $\delta$ and
${\cal N}$.  For the energy shift $\delta$ we have used two forms for
the prior, either a top hat spanning the energy-scale uncertainty of
the experiment, $\sigma_{E_s}$, or a gaussian prior of width
$\sigma_{E_s}$.

For ${\cal N}$ we impose the prior arising from requiring consistency
with the Fermi-LAT measurements \cite{Abdo:2010nz} of the diffuse
extra-galactic $\gamma$-ray background. In order to do so we obtain the
total energy density of EM radiation from the proton propagation using
Eq.~\eqref{eq:omegacas} and we require following
Ref.~\cite{Berezinsky:2010xa}:
\begin{equation}
w_{\rm cas}({\cal N} ,n,\gamma)\leq 5.8\times 10^{-7}\; {\rm eV}/{\rm cm}^3\,. 
\label{eq:fermilat}
\end{equation}

The marginalization in Eq.~(\ref{eq:margi}) also determines ${\cal
  N}_{\rm best}$ and $\delta_{\rm best}$ for the model, which are the
values of the energy shift and normalization that yield the best
description of the experimental CR data, subject to the constraint
imposed by the Fermi-LAT measurement.

Altogether the model is compatible with the experimental results at
given goodness of the fit (GOF) if
\begin{equation}
\sum_{\vec k}P_{\vec k}(n,\gamma,{\cal N}_{\rm best},\delta_{\rm best})
\Theta\left[P_{\vec k}(n,\gamma,{\cal N}_{\rm best},\delta_{\rm best})
-P_{\rm exp}(n,\gamma)\right]\leq {\rm GOF}
\end{equation}
Technically, this is computed by generating a large number $N_{\rm
  rep}$ of replica experiments according to the probability
distribution $P_{\vec k}(n,\gamma,{\cal N}_{\rm best},\delta_{\rm
  best})$ and imposing the fraction $F$ of those which satisfy $P_{\vec
  k}(n,\gamma,{\cal N},\delta_{\rm best})>P_{\rm exp}(n,\gamma)$ to be 
$F\leq{\rm GOF}$.

\begin{table}[t]\centering
\begin{minipage}[t]{0.7\linewidth}\centering\small
\setlength{\tabcolsep}{3pt}\renewcommand{\arraystretch}{1.3}\small
\begin{tabular}{c||cc|c|cc||cc|c|cc}
\hline\hline 
&\multicolumn{5}{c||}{\normalsize $E_{\rm min}=10^{17.5}$~eV}
&\multicolumn{5}{c}{\normalsize $E_{\rm min}=10^{18}$~eV}\\
model&$n$&$\gamma$
&$\omega_{\rm cas}$${}^a$\footnotetext[1]
{in units of $10^{-7}$~eV/cm${}^3$}
&$\delta_{I \rm best}$&$\delta_{II\rm best}$
&$n$&$\gamma$&$\omega_{\rm cas}$${}^a$ 
&$\delta_{I\rm best}$&$\delta_{II\rm best}$
\\
\hline &\multicolumn{10}{c}{fit {\it with} Fermi-LAT bound:}\\ \hline
best fit
&$3.50$&$2.49$&$5.8$&0.005&0.
&$3.20$&$2.52$&$5.2$& 0.050&0.045
\\
min.~$\omega_{\rm cas}$
&$4.50$&$2.31$&$4.4$&-0.235&-0.245
&$2.25$&$2.47$&$1.7$&-0.120&-0.150 
\\
max.~$\omega_{\rm cas}$
&$4.60$&$2.36$&$5.8$ &-0.185&-0.175
&$3.35$&$2.55$&$5.8$ &0.050&0.060
\\
min.~$\omega_\pi$
&$2.00$&$2.67$&$4.9$&0.215&0.235 
&$2.00$&$2.51$&$1.8$& -0.070&-0.095
\\
max.~$\omega_\pi$
&$4.80$&$2.29$&$5.8$&-0.220&-0.215
&$5.10$&$2.29$&$5.8$&-0.250&-0.250
\\
\hline &\multicolumn{10}{c}{fit {\it without} Fermi-LAT bound:}\\ 
\hline
max.~$\omega_{\rm cas}$
&$4.45$&$2.44$&$15$ &0.135&0.155 
&$5.25$&$2.36$&$27$ & 0.205&0.205
\\
max.~$\omega_\pi$
&$4.80$&$2.36$&$14$& 0.050&0.055
&$5.30$&$2.35$&$26$& 0.190&0.190
\\\hline\hline
\multicolumn{11}{c}{}\\
\hline\hline 
&\multicolumn{5}{c||}{\normalsize $E_{\rm min}=10^{18.5}$~eV}
&\multicolumn{5}{c}{\normalsize $E_{\rm min}=10^{19}$~eV}\\
model
&$n$&$\gamma$ &$\omega_{\rm cas}$${}^a$
&$\delta_{I \rm best}$&$\delta_{II\rm best}$
&$n$&$\gamma$&$\omega_{\rm cas}$${}^a$ 
&$\delta_{I\rm best}$&$\delta_{II\rm best}$
\\
\hline &\multicolumn{10}{c}{fit {\it with} Fermi-LAT bound:}\\ \hline
best fit
&$4.05$&$2.47$&$5.8$& 0.015&0.005
&$4.60$&$2.50$&$4.4$& -0.030&-0.065
\\
min.~$\omega_{\rm cas}$
&$2.00$&$2.45$&$1.4$ & -0.050&-0.060 
&$2.00$&$2.88$&$0.44$ &-0.220&-0.250
\\
max.~$\omega_{\rm cas}$
&$4.95$&$2.37$&$5.8$ & -0.165&-0.160
&$4.45$&$2.13$&$5.8$ & 0.130&0.090
\\
min.~$\omega_\pi$
&$2.00$&$2.63$&$2.1$& 0.075&0.070
&$2.00$&$2.88$&0.44 & -0.220&-0.250
\\
max.~$\omega_\pi$
&$5.35$&$2.28$&$5.8$&-0.240&-0.250
&$4.40$&$2.10$&$5.8$&0.145&0.100
\\
\hline &\multicolumn{10}{c}{fit {\it without} Fermi-LAT bound:}\\ 
\hline
max.~$\omega_{\rm cas}$
&$6.00$&$2.49$&$30$ & 0.120&0.135
&$6.00$&$2.14$&$23$ &0.250&0.210
\\
max.~$\omega_\pi$
&$6.00$&$2.47$&$29$& 0.120&0.125
&$6.00$&$2.10$&$23$&0.250&0.210 \\
\hline\hline
\end{tabular}
\end{minipage}
\caption[]{Cosmic ray source parameters which best fit the HiRes 
  data~\cite{Abbasi:2007sv}, along with those which yield minimal and maximal 
  contributions to $\omega_\pi$ (i.e. neutrino fluxes) and 
  $\omega_\mathrm{cas} = \omega_\pi + \omega_\mathrm{BH}$ (i.e. 
  $\gamma$-ray fluxes), all at the 99\% C.L.}
\label{tab:parameters}
\end{table}

\begin{figure}[t]
\begin{center}
\includegraphics[width=\linewidth]{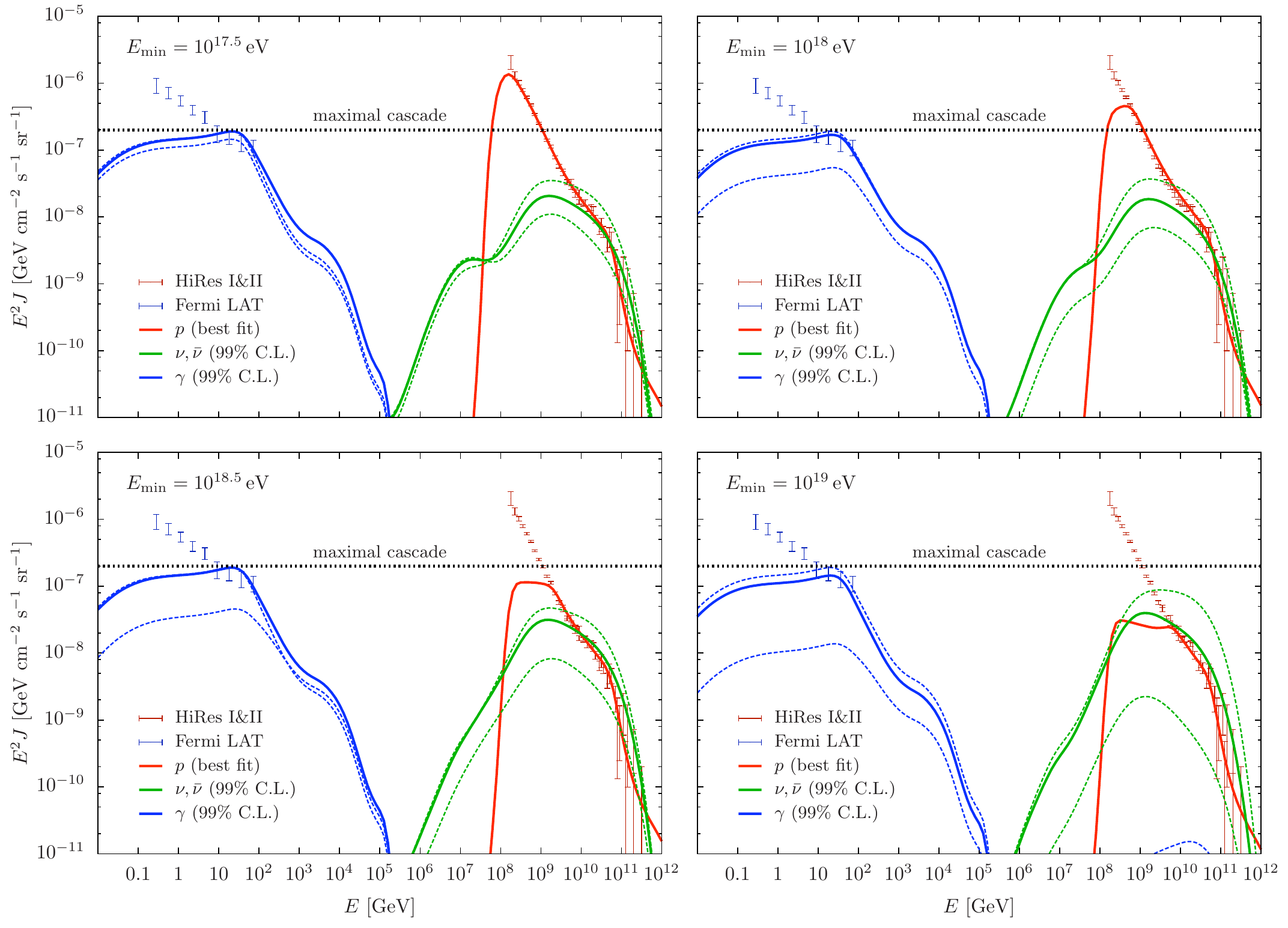}
\end{center}
\vspace{-0.3cm}
\caption[]{Comparison of proton, neutrino and $\gamma$-ray fluxes for
  different crossover energies. We show the best-fit values (solid
  lines) as well as neutrino and $\gamma$-ray fluxes within the 99\%
  C.L. with minimal and maximal energy density (dashed lines). The
  values of the corresponding model parameters can be found in
  Table.~\ref{tab:parameters}. The dotted line labeled ``maximal
  cascade'' indicates the approximate limit $E^2 J_{\rm cas} \lesssim
  c\, \omega_{\rm cas}^{\rm max}/4\pi\log({\rm TeV}/{\rm GeV})$,
  corresponding to a $\gamma$-ray flux in the GeV-TeV range saturating
  the energy density~(\ref{eq:fermilat}). The $\gamma$-ray fluxes are
  marginally consistent at the 99\% C.L. with the highest energy
  measurements by Fermi-LAT. The contribution around 100~GeV is
  somewhat uncertain due to uncertainties in the cosmic infrared
  background.}
\label{fig:comparison}
\end{figure}

\begin{figure}[t]
\begin{center}
\includegraphics[width=\linewidth]{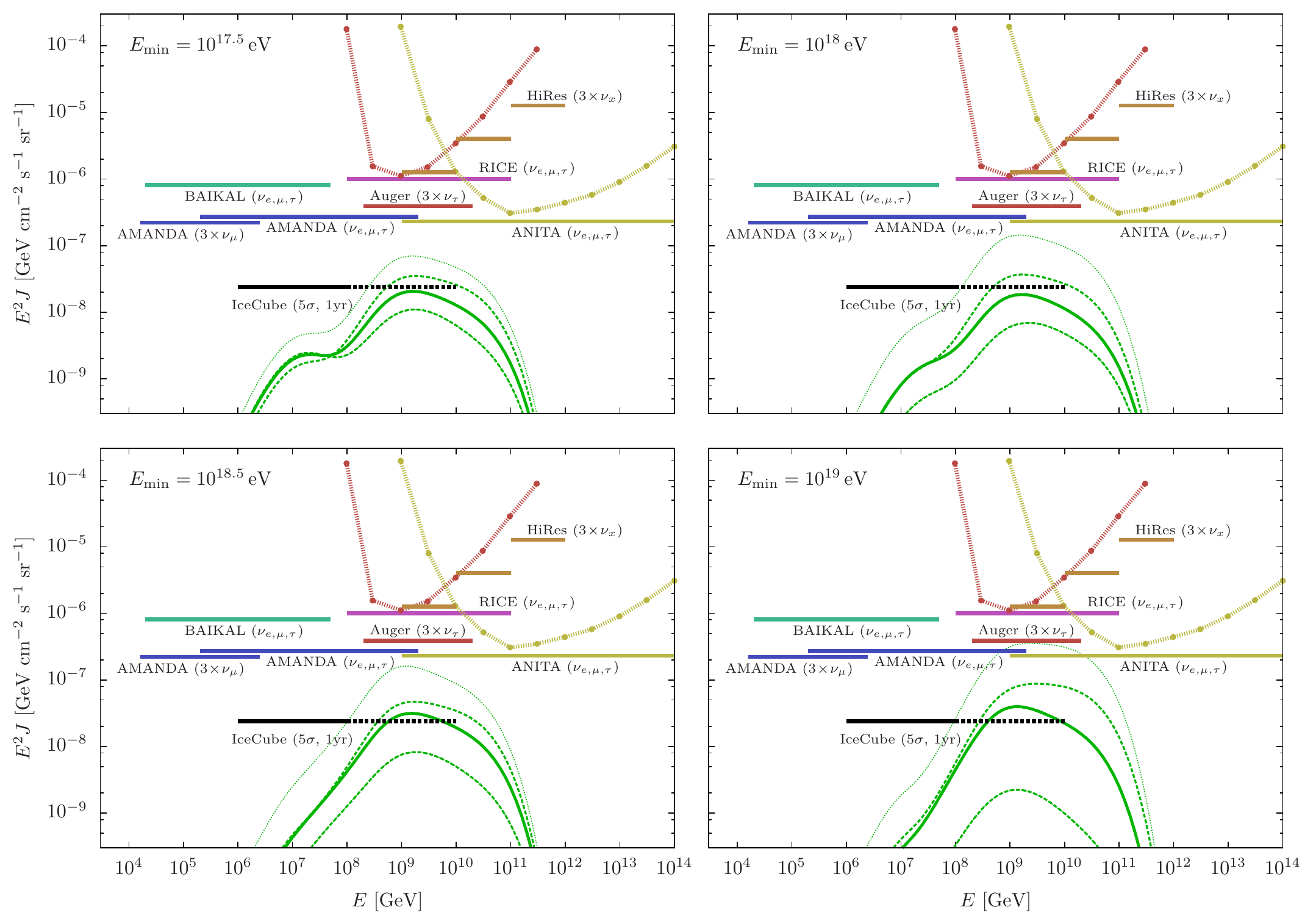}
\end{center}
\vspace{-0.3cm}
\caption[]{The predicted best fit (solid) and 99\% C.L. range of
  cosmogenic neutrino fluxes with (dashed) and without (dotted) the
  Fermi-LAT constraint. The values of the corresponding model
  parameters can be found in Table.~\ref{tab:parameters}. For
  comparison we show upper limits on the total diffuse neutrino flux
  from AMANDA~\cite{Achterberg:2007qp,Ackermann:2007km},
  Auger~\cite{Collaboration:2009uy}, Lake
  Baikal~\cite{Aynutdinov:2005dq}, HiRes~\cite{Martens:2007ff}
  (minimum of $\nu_\mu$ and $\nu_\tau$ channel),
  RICE~\cite{Kravchenko:2006qc} and ANITA~\cite{Barwick:2005hn}. The
  black solid line shows the 5$\sigma$ sensitivity of IceCube after
  just 1 year of observation \cite{Achterberg:2006md}. The cutoff at
  $10^8$ GeV is artificial so we also show an extrapolation to higher
  energies as a black dashed line following
  Ref.\cite{Halzen:2006ic}). All limits are obtained assuming an equal
  distribution between neutrino flavours:
  $N_{\nu_e}:N_{\nu_\mu}:N_{\nu_\tau}\sim 1:1:1$ (and scaled
  appropriately where necessary). Integrated limits assuming an
  $E^{-2}$ spectrum are shown as solid lines and differential limits
  as dotted lines (both limits are shown for Auger and ANITA).}
\label{fig:nulimits}
\end{figure}

With this method we determine the value of $(n, \gamma)$ parameters
that are compatible with the HiRes I and HiRes II
experiments~\cite{Abbasi:2007sv}.  We plot in the left panel of
Fig.~\ref{fig:gof} the regions with GOF 64\%, 95\% and 99\% for four
values of the minimum ({\em i.e.} crossover) energy. In the right
panel we show the corresponding ranges of $w_{\rm cas,best}$ for the
models as a function of the cosmic evolution index $n$. In order to
display explicitly the impact of the constraint from the Fermi-LAT
measurements of the diffuse extra-galactic $\gamma$-ray background
\eqref{eq:fermilat}, we show the corresponding GOF regions without
imposing that constraint.  In Table.~\ref{tab:parameters} we list the
parameters corresponding to the best-fit models and to the models with
minimal and maximal contributions to $\omega_\pi$ and $\omega_{\rm
  cas}=\omega_\pi+\omega_{\rm BH}$ at the 99\% C.L., together with the
corresponding energy shifts which give best fits to the HiRes I and
Hires II data.  We also show the parameters for the models with
maximum $\omega_\pi$ and $\omega_{\rm cas}$ {\em without} imposition
of the Fermi-LAT constraint.

As an illustration of the agreement with the CR data we show in
Fig.~\ref{fig:gof} the range of proton fluxes corresponding to models
with GOF 99\% or better for increasing crossover energies $E_{\rm
  min}$. As discussed above each fit of the proton spectra is
marginalized with respect to the experimental energy scale uncertainty
and we show the shifted predictions with $\delta_{\rm best}$ in
comparison to the HiRes data at central value.  We also show in the
figure the results from Auger~\cite{Abraham:2008ru,Abraham:2010mj},
though these have not been included in the analysis (see below).

These results are obtained assuming an energy scale uncertainty
$\sigma_{E_s}=5\%$ with a ``top-hat'' prior for the corresponding
energy shifts which are taken to be uncorrelated for HiRes I and HiRes
II.  In Fig.~\ref{fig:compa} we explore the dependence of the results
on these assumptions by using a different form for the prior, assuming
the energy shifts to be correlated between the two experiments, or
reducing the uncertainty to $\sigma_{E_s}=15\%$. As seen in the
figure, the main effect is associated with the reduction of the energy
scale uncertainty which, as expected, results in a worsening of the
GOF for models with larger $n$. This is directly related to the
normalization constraint from Eq.~(\ref{eq:fermilat}).  If one naively
ignores the energy scale uncertainty, the constraint in
Eq.~(\ref{eq:fermilat}) rules out models with $n\gtrsim 3$ (the
precise value depending on the assumed $E_{\rm min}$). However, once
the energy scale uncertainty is included, the constraint of
Eq.~(\ref{eq:fermilat}) plays a weaker role on the determination of
the GOF of the models. It does however imply a maximum value of ${\cal
  N}_{\rm best}$ which, as we will see, impacts the corresponding
ranges of neutrino fluxes.

The corresponding range of $\gamma$-ray and cosmogenic neutrino fluxes
(summed over flavour) is shown in Fig.~\ref{fig:comparison} for models
with minimal and maximal energy density at the 99\% C.L.
As expected, the maximum $\gamma$-ray fluxes are consistent with the
Fermi-LAT data within the errors.  For illustration, we also show as a
dotted line the ``naive'' $\gamma$-ray limit $E^2J_{\rm cas} \lesssim
c\,\omega_{\rm cas}^{\rm max}/4\pi\log({\rm TeV}/{\rm GeV})$,
corresponding to a $\gamma$-ray flux in the GeV-TeV range which saturates the
energy density~(\ref{eq:fermilat}).

We have not included in the analysis the results from the Auger
Collaboration \cite{Abraham:2008ru,Abraham:2010mj},  which are shown
in Fig.\ref{fig:omega} for illustration only (hence our results are
directly comparable to those in Ref.\cite{Berezinsky:2010xa}).  As
described in Refs.~\cite{Abraham:2008ru,Abraham:2010mj}, besides the
energy scale uncertainty there is also an (energy-dependent) energy
resolution uncertainty which implies that bin-to-bin migrations
influence the reconstruction of the flux and spectral shape. Since the
form of the corresponding error matrix is not public, this data
\cite{Abraham:2008ru,Abraham:2010mj} cannot be analysed outside the
Auger Collaboration.

\section{Discussion}\label{sec:III}

The cosmogenic neutrino fluxes that we have shown in
Fig.~\ref{fig:comparison} are compared to present upper limits on the
diffuse neutrino flux in Fig.~\ref{fig:nulimits}. As before, the solid
green line shows the neutrino flux (summed over flavours)
corresponding to the best fit of the proton spectra and the dashed
green line indicate the range of neutrino fluxes within the
99\%~C.L. For all crossover energies considered, the range of models
at the 99\% C.L.~is consistent with existing neutrino limits.
For illustration, the thin dotted line shows the larger range of neutrino
fluxes at the 99\% C.L.~corresponding to a fit {\em without} the Fermi
LAT constraint ({\it cf.}~the black contours in the left panel of
Fig.~\ref{fig:gof}). It is apparent that this indirect bound from
GeV-TeV $\gamma$-rays does reduce the number of possible models
significantly.

At this point it is worth stressing that the Fermi-LAT spectrum used 
in this analysis is not the result of a direct observation but is 
derived by a foreground subtraction scheme. The extra-galactic $\gamma$-ray 
background inferred by EGRET~\cite{Sreekumar:1997un} shows a 
significantly larger intensity and a harder spectral index. A possible 
source of the differences could be due to the different diffuse galactic 
emission (DGE) models used in the analysis. As pointed out in~\cite{Abdo:2010nz} 
a re-analysis of the EGRET data with an updated DGE model~\cite{Strong:2004ry} 
is comparable with the intensity observed with Fermi-LAT. It is beyond 
the scope of this paper to address these systematic uncertainties. 
The {\it maximal} effect of a larger $\gamma$-ray background intensity is 
indicated by the extended parameter regions shown in Fig.~\ref{fig:gof} 
which are derived {\em without} the Fermi-LAT constraint together with the 
corresponding range of neutrino fluxes in Fig.~\ref{fig:nulimits}.

The overall range of neutrino fluxes increases along with the
crossover energy - not only in magnitude, which is expected already
due to the reduced set of CR data used in the GOF test, but also to
significantly larger neutrino fluxes. Also the cosmogenic neutrino
flux of the best-fit models increases by over a factor of two in the
peak region ($\sim 10^9$~GeV). This confirms our earlier suspicion
that an increasing value of the crossover energy allows a larger
contribution of cosmogenic neutrinos relative to the $\gamma$-rays and
hence larger neutrino fluxes.

Figure~\ref{fig:nulimits} also shows the estimated sensitivity of
IceCube~\cite{Ahrens:2003ix} (5$\sigma$) to neutrino fluxes in the
$10^6$-$10^8$~GeV (solid)~\cite{Achterberg:2006md} and the
$10^8$-$10^{10}$~GeV (dotted)~\cite{Halzen:2006ic} energy range after
one year of observation. IceCube located at the South Pole is
presently the largest neutrino telescope. On completion in early 2011
it will consist of a km${}^3$-scale detector of transparent glacial
ice, that is constantly monitored for \v{C}erenkov light emission of
secondary charged particles from high energy neutrino interactions. It
is apparent from Fig.~\ref{fig:nulimits} that IceCube's sensitivity
after one year is already sufficient to probe cosmogenic neutrino
fluxes from an all-proton spectrum of extra-galactic cosmic rays. If
the crossover energy exceeds $10^{18.5}$~eV, the best-fit model of the
HiRes data is within reach of IceCube.

In summary we find that while the expected range of cosmogenic
neutrino fluxes in all-proton models is indeed reduced due to the
constraint from the diffuse $\gamma$-ray background measurements,
neutrino fluxes compatible at 99\% C.L with the HiRes and Fermi-LAT
results can be larger than those presented in
Ref.\cite{Berezinsky:2010xa} by up to factor of $\sim 30$ for the same
values of $E_{\rm max}=10^{21}$ eV and $z_{\rm max}=2$. In particular
the allowed cosmogenic flux is still within reach of neutrino
observatories like IceCube.  Furthermore, 
our results are obtained with the simple parametrization of the source
spectral emission rate in Eq.~\eqref{eq:injection}; larger
neutrino fluxes might be allowed with a more general spectrum
than a simple power-law.

One can also turn this argument around and use observation or
non-observation of cosmogenic neutrinos in the near future to provide
additional constraints on the composition of cosmic
rays~\cite{Ahlers:2009rf}. We have assumed here an
all-proton composition for extra-galactic cosmic rays. However, as mentioned already,
the chemical composition of UHE CRs is rather uncertain and may well
be dominated by heavy nuclei. In this case the limits on diffuse
neutrino fluxes can still serve as a probe of the possible proton
fraction in cosmic rays~\cite{Ahlers:2009rf,Anchordoqui:2009nf} and
the limits from diffuse $\gamma$-rays serve as an additional
probe~\cite{Ahlersetal}.

\section*{Acknowledgments}
The authors would like to thank Andrew Taylor and the anonymous referee for 
helpful comments on the manuscript.
This work is supported by US National Science Foundation Grant No
PHY-0757598 and PHY-0653342, by the Research Foundation of SUNY at
Stony Brook, the UWM Research Growth Initiative. F.H.~is supported by
U.S. National Science Foundation-Office of Polar Program,
U.S. National Science Foundation-Physics Division, and the University
of Wisconsin Alumni Research Foundation. M.C.G.-G.~acknowledges further
support from Spanish MICCIN grants 2007-66665-C02-01, ACI2009-1038,
consolider-ingenio 2010 grant CSD2008-0037 and by CUR Generalitat de
Catalunya grant 2009SGR502. S.S.~acknowledges support by the EU Marie 
Curie Network ``UniverseNet'' (HPRN-CT-2006-035863).

\appendix

\section{Cascade Solution}\label{app:I}

The relevant processes with background photons contributing to the
differential interaction rates $\gamma_{ee}$, $\gamma_{\gamma e}$ and
$\gamma_{e \gamma}$ are inverse Compton scattering (ICS),
$e^\pm+\gamma_{\rm bgr}\to e^\pm+\gamma$, pair production (PP),
$\gamma+\gamma_{\rm bgr}\to e^++e^-$, double pair production (DPP)
$\gamma+\gamma_{\rm bgr}\to e^++e^-+e^++e^-$, and triple pair
production (TPP), $e^\pm+\gamma_{\rm bgr}\to
e^\pm+e^++e^-$~\cite{Blumenthal:1970nn,Blumenthal:1970gc}.  The
angular-averaged (differential) interaction rate, $\Gamma_i$
($\gamma_{ij}$) is defined as
\begin{gather}\label{Gamma}
\Gamma_{i}(z,E_i) =
\frac{1}{2}\int\limits_{-1}^1\mathrm{d}\cos\theta\int\mathrm{d}
\epsilon\,(1-\beta
\cos\theta) n_\gamma(z,\epsilon)\sigma^\mathrm{tot}_{i\gamma}\,,\\
\gamma_{ij}(z,E_i,E_j) = \Gamma_i(z,E_i)\,\frac{\D N_{ij}}{\D E_j}(E_i,E_j)\,,
\label{gamma}
\end{gather}
where $n_\gamma(z,\epsilon)$ is the energy distribution of background
photons at redshift $z$ and $\D N_{ij}/\D E_j$ is the angular-averaged
distribution of particles $j$ after interaction of particle $i$.

\begin{figure}[t]
\begin{center}
\includegraphics[width=0.6\linewidth]{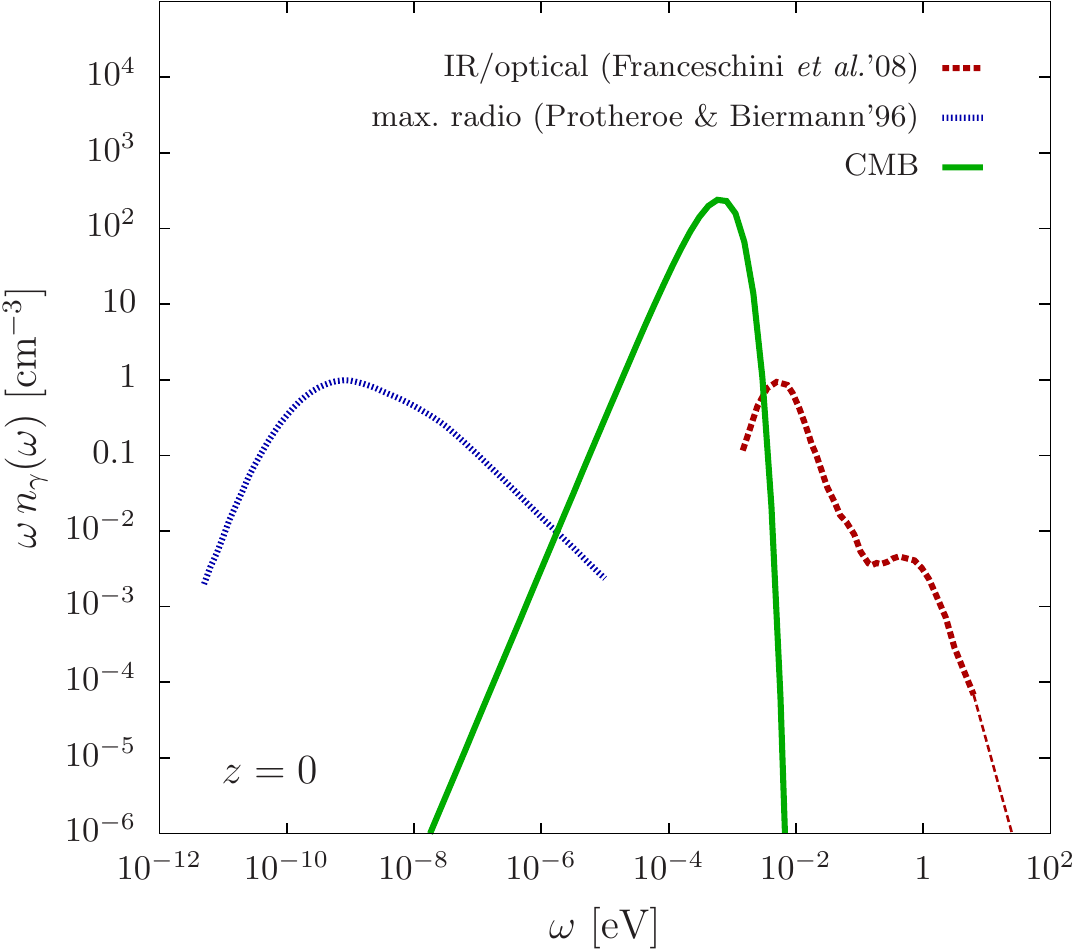}
\end{center}
\vspace{-0.5cm}
\caption[]{The energy spectrum of the CMB~\cite{Amsler:2008zzb} and
  the CIB in the IR/optial~\cite{Franceschini:2008tp} and
  radio~\cite{Protheroe:1996si} range at $z=0$. The thin dashed line 
  shows our extrapolation to UV energies.}
\label{fig:CRB}
\end{figure}

Besides the contribution of the CMB the shape of the cascade spectrum
depends on the cosmic infrared/optical background. 
We will use a recent estimate~\cite{Franceschini:2008tp} (that we extrapolate 
slightly to UV energies as seen in Fig.~\ref{fig:CRB}) and assume a 
redshift dependence following the star formation rate as described in 
Ref.~\cite{Ahlers:2009rf}. This is consistent with the constraints on
the $\gamma$-ray opacity of the universe set by HESS
\cite{Aharonian:2005gh}, MAGIC \cite{Aliu:2008ay} and Fermi-LAT
\cite{LAT:2010kz}.

We have little direct knowledge of the cosmic radio background. An
estimate made using the RAE satellite~\cite{Clark} is often used to
calculate the cascading of UHE photons~\cite{Bhattacharjee:1998qc}. A
theoretical estimate has been made~\cite{Protheroe:1996si} of the
intensity down to kHz frequencies, based on the observed luminosity
function and radio spectra of normal galaxies and radio galaxies
although there are large uncertainties in the assumed evolution. The
calculated values are about a factor of $\sim 2$ above the
measurements and to ensure maximal energy transfer in the cascade we
will adopt this estimate and assume the same redshift scaling as the
cosmic infrared/optical background. The magnitude of the adopted radio
background radio is not important for the shape of the GeV-TeV
spectrum as can be seen from Fig.~6 of Ref.~\cite{Sarkar:2003sp} where
even higher values are considered. We summarize the adopted cosmic
radiation backgrounds in Fig.~\ref{fig:CRB}.

High energetic electrons and positrons may also lose energy via
synchrotron radiation in the intergalactic magnetic field $B$ with a
random orientation $\sin\theta$ with respect to the velocity
vector. We will assume in the following that the field strength $B$
is of ${\cal O}(10^{-12})$G~\cite{Dolag:2004kp},
which leads to an efficient transfer of energy into the EM
cascade. The synchrotron power spectrum (W eV${}^{-1}$) has the form
\begin{equation}
\mathcal{P}(E_e,E_\gamma) = \frac{\sqrt{3}\alpha}{2\pi}
\frac{eB\sin\theta}{m_e}F(E_\gamma/E_{\rm c})\,;\qquad F(t) \equiv
t\int_t^\infty{\rm d}z\, K_{5/3}(z)\,,
\end{equation}
where we follow the notation of Ref.~\cite{Schlickeiser:2002pg} with
$E_{\rm c} = (3eB\sin\theta/2m_e)(E_e/m_e)^2$.
This can be treated as a continuous energy loss of the electrons and
positrons with a parameter\footnote{Note, the identity $\int{\rm
    d}E\left[E\,\partial_E(bn_e) + \int{\rm d}E'
    \mathcal{P}(E',E)n_e\right] = 0$, implying overall energy
  conservation.}
\begin{equation}
b_{\rm syn}(E_e) = \frac{1}{2}\int{\rm d}\cos\theta\int{\rm
d}E_\gamma\mathcal{P}(E_e,E_\gamma) =
\frac{4\alpha}{9}\left(\frac{eB}{m_e}\right)^2\left(\frac{E_e}{m_e}\right)^2\,.
\end{equation}
We will assume in the calculation that the intergalactic magnetic
field is primordial with a (flux-conserving) redshift dependence $B(z) =
(1+z)^2B(0)$. Note, that the synchrotron energy loss has then a
redshift dependence similar to BH pair production in the CMB, {\it
  i.e.}~$b_{\rm syn}(z,E) = (1+z)^2b_{\rm syn}(0,(1+z)E)$.  It is also
convenient to define $\gamma_{e\gamma}^{\rm syn}(E_e,E_\gamma) \equiv
\mathcal{P}(E_e,E_\gamma)/E_\gamma$, which has an analogous redshift
dependence\, {\it i.e.}~$\gamma_{e\gamma}^{\rm syn}(z,E_e,E_\gamma) =
(1+z)^4\gamma_{e\gamma}^{\rm syn}(0,(1+z)E_e,(1+z)E_\gamma)$.

The fast evolution of the cascade is governed by the set of
differential equations,
\begin{align}\label{eq:CAS1}
\partial_{\hat{t}}Y_\gamma(E) =&-\Gamma_\gamma(E)Y_\gamma(E) +
\int{\rm d}E'\frac{\mathcal{P}(E',E)}{E}Y_e(E')+\int{\rm d}
E'\gamma_{e\gamma}(E',E)Y_e(E')\,,\\ \partial_{\hat{t}}Y_e(E)
=&-\Gamma_e(E)Y_e(E)+\partial_E(b(E)Y_e(E)) + \int{\rm d}
E'\left[\gamma_{\gamma
e}(E',E)Y_\gamma(E')+\gamma_{ee}(E',E)Y_e(E')\right]\,,\label{eq:CAS2}
\end{align}
which determines the evolution on {\it short} time-scales
$\Delta\hat{t}\,\Gamma_{p\gamma}\ll1$ (the redshift $z$ is
kept {\it fixed} meanwhile). The initial
condition $Y_{\gamma/e}(E)|_{\hat t=0}$ is given by the sum of
previously developed cascades and the newly generated contributions
from proton interactions.

The solution of Eqs.~(\ref{eq:CAS1}) and (\ref{eq:CAS2}) for an
infinitesimally small step $\Delta \hat t$ can be written for a
discrete energy spectrum, $N_i \simeq \Delta E_iY_i$, as
\begin{equation}\label{eq:discrete}
\begin{pmatrix}N_\gamma\\N_e\end{pmatrix}_i(\hat{t}+\Delta\hat{t})
  \simeq \sum_j\begin{pmatrix}T_{\gamma\gamma}(\Delta\hat
  t)&T_{e\gamma}(\Delta\hat t)\\T_{\gamma e}(\Delta\hat
  t)&T_{ee}(\Delta\hat
  t)\end{pmatrix}_{ji}\begin{pmatrix}N_\gamma\\N_e\end{pmatrix}_j(\hat{t})\,.
    \end{equation}
    With the transition matrix $\mathcal{T}(\Delta \hat t)$, defined
    by Eq.~(\ref{eq:discrete}), we can efficiently follow the
    development of the EM cascade over a distance $\Delta
    t = 2^N\Delta \hat t$ via matrix doubling~\cite{Protheroe:1992dx}:
\begin{equation}
\mathcal{T}(2^N\Delta\hat t) \simeq \left[\mathcal{T}(\Delta\hat
t)\right]^{N+1}\,.
\end{equation}

We will compare our calculation with results from other investigations
in Appendix \ref{comp}.

\section{Energy Density of the Cascade}\label{app:II}

We can express the system of partial integro-differential
equations~(\ref{diff0}) as,
\begin{align}\label{diff1}
  \dot Z_i &=
  \partial_{\mathcal{E}}\left(b_i(z,\mathcal{E})Z_i(z,E)\right)-\Gamma_{i}(z,\mathcal{E})\,Z_i+(1+z)\mathcal{L}^\mathrm{eff}_i(z,\mathcal{E})\,,
\end{align}
where we have defined $\mathcal{E} = (1+z)E$, and $Z_i(z,E) \equiv
(1+z)Y_i(z,\mathcal{E})$, subject to the boundary condition
$Z_j(z_{\rm max},E) = 0$. The effective source term in
Eq.~(\ref{diff1}) is
\begin{equation}
  \mathcal{L}^{\rm eff}_i =\mathcal{L}_i +\sum_j\int{\rm d}
  E_j\,\gamma_{ji}(z,\mathcal{E}_j,E_i)\,Z_j\,,\label{Leff}
\end{equation}
The total energy of the cascade is given in
Eq.~(\ref{eq:omegacas}). This can be obtained by integrating
Eq.~(\ref{diff1}):
\begin{equation}
\label{eq:DE}
\frac{{\rm d}}{{\rm d} t}\left[\int{\rm d} E E Z_{\rm cas}(z,E)\right]
= - \int{\rm d} E E\partial_{\mathcal{E}}\left[b_{\rm
cas}(z,\mathcal{E})Z_{\rm p}(z,E)\right]\,.
\end{equation} 
Integrating the r.h.s.~by parts yields
\begin{equation}
{\rm r.h.s.} = - \int{\rm d} E\partial_{E}\left[ E\frac{1}{1+z}b_{\rm
cas}(z,\mathcal{E})Z_{\rm p}(z,E)\right] + \int{\rm d} E
\frac{1}{1+z}b_{\rm cas}(z,\mathcal{E})Z_{\rm p}(z,E)\,.
\end{equation}
The first term vanishes since $b_{\rm cas}=0$ for sufficiently low
energies and $Z_{\rm p}=0$ beyond the maximal energy. The time
integration of the l.h.s.~between the present epoch ($t=0$) and the
first sources ($t_{\rm max}$) gives
\begin{equation}
\int_0^{t_{\rm max}}{\rm d}t [{\rm l.h.s.}] = \int{\rm d} E E n_{\rm
cas}(E) = \omega_{\rm cas}\,,
\end{equation}
hence we obtain Eq.~(\ref{eq:omegacas}).

\begin{figure}[t]
\begin{center}
\includegraphics[width=0.5\linewidth]{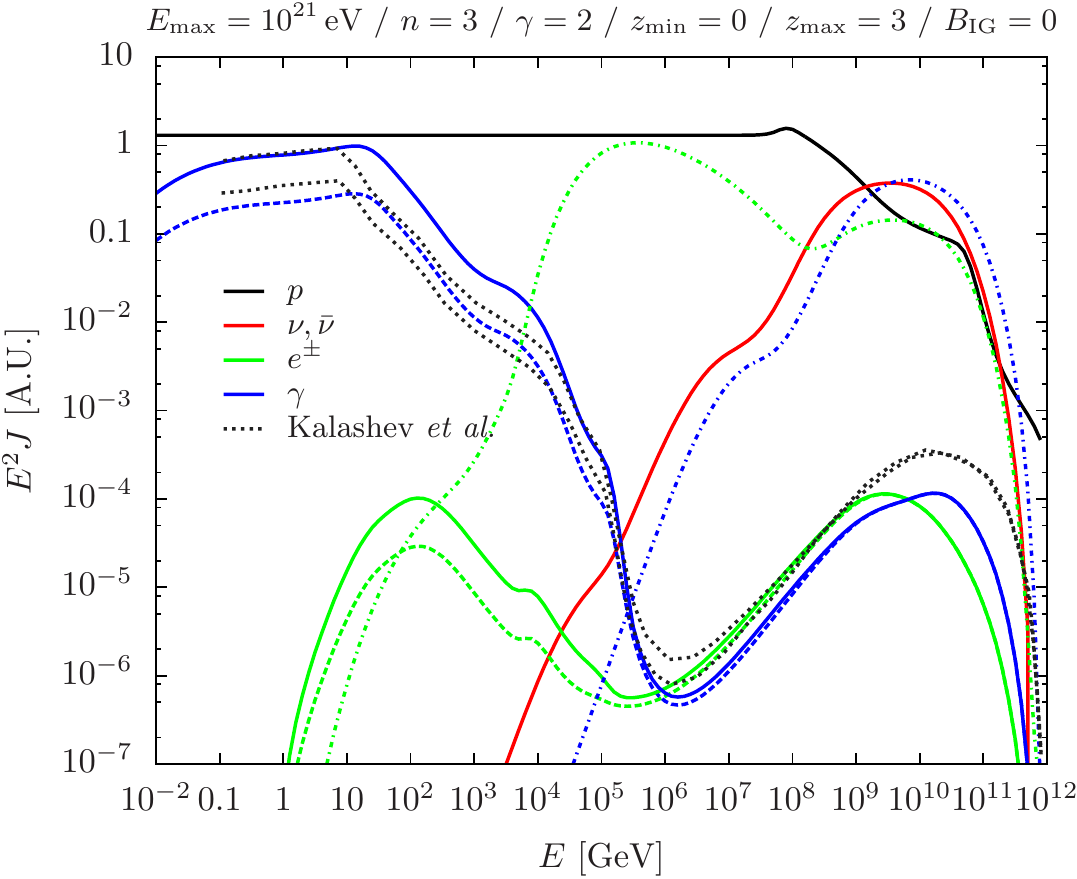}\hfill
\includegraphics[width=0.5 \linewidth]{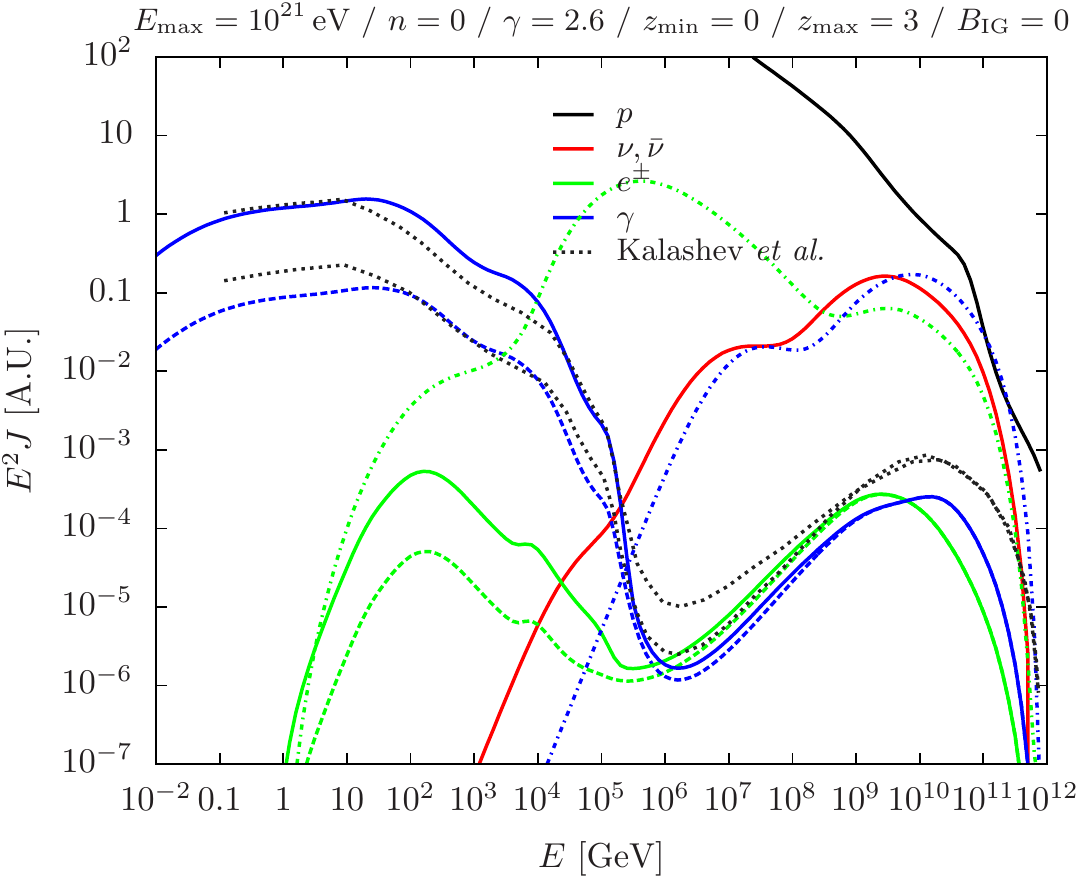}
\end{center}
\vspace{-0.5cm}
\caption[]{Comparison of our calculations with the $\gamma$-ray spectra of
Ref.~\cite{Kalashev:2007sn} shown as black doted lines. We normalize
the $\gamma$-ray (green) and electron/positron (red) spectra to the
proton spectra (black solid line). For comparison, the dashed lines
show the spectra without the contribution of BH pairs and
the dashed-dotted lines show the spectra without EM
cascades.}\label{fig:compgamma}
\end{figure}

\section{Comparison of Gamma-Ray Spectra}\label{comp}

Figure~\ref{fig:compgamma} compares our calculations with
spectra derived in Ref.~\cite{Kalashev:2007sn} (Figs.~1 and 2) using
$n=3$ and $\gamma=2$ (left plot) as well as $n=0$ and $\gamma=2.6$
(right plot), respectively, with (upper) and without (lower) BH
contributions.

The $\gamma$-ray spectra without BH pairs in the cascade are consistent
with Ref.~\cite{Kalashev:2007sn} apart from a slightly
smaller energy density (recognizable as an overall shift downwards) probably due
to the difference in the adopted IR/optical background. Since the energy density derived from our spectra agrees
with the value obtained from Eq.~(\ref{eq:omegacas}) within 10\%, we
believe that the overall normalization of our spectra is correct.

For a full calculation, {\it i.e.}~including the BH pairs, our overall
energy density agrees well with the calculation in
Ref.~\cite{Kalashev:2007sn}. Again, the energy density derived from
our spectra agrees with the value obtained from
Eq.~(\ref{eq:omegacas}) within 10\%. The pair production dip at
$10^6$~GeV is more pronounced in our spectra. Note that the energy
loss of the cascade beyond $10^5$ GeV is much more rapid than the
Bethe-Heitler pair production rate, hence the modest increase in
$\gamma$-rays beyond this energy meets our expectations. Moreover,
instead of using a power-law approximation
($\mathrm{d}n/\mathrm{d}E_{\pm} \propto E_{\pm}^{-7/4}$) for the BH
$e^\pm$ spectrum (see Ref.\cite{Kelner:2008ke} for a critical
discussion) we use the exact differential cross-section of
Ref.\cite{Blumenthal:1970nn}.

\begin{figure}[t]
\begin{center}
\includegraphics[width=0.5\linewidth]{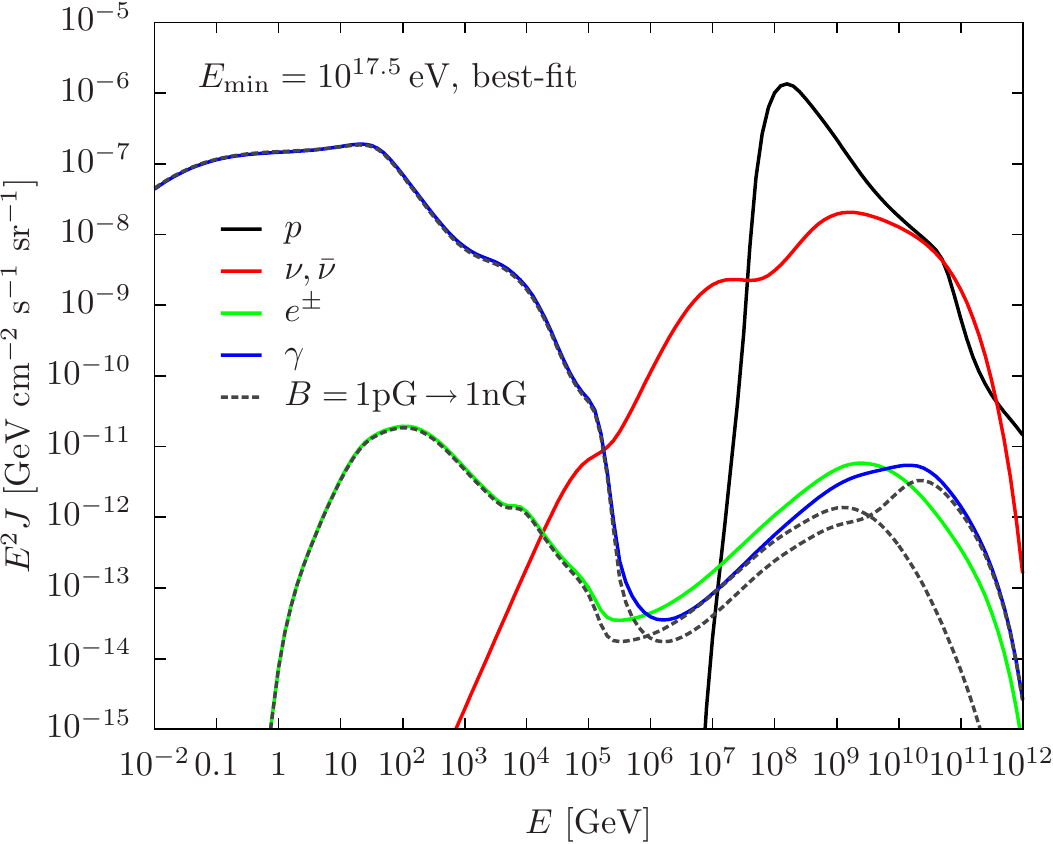}\hfill
\includegraphics[width=0.5 \linewidth]{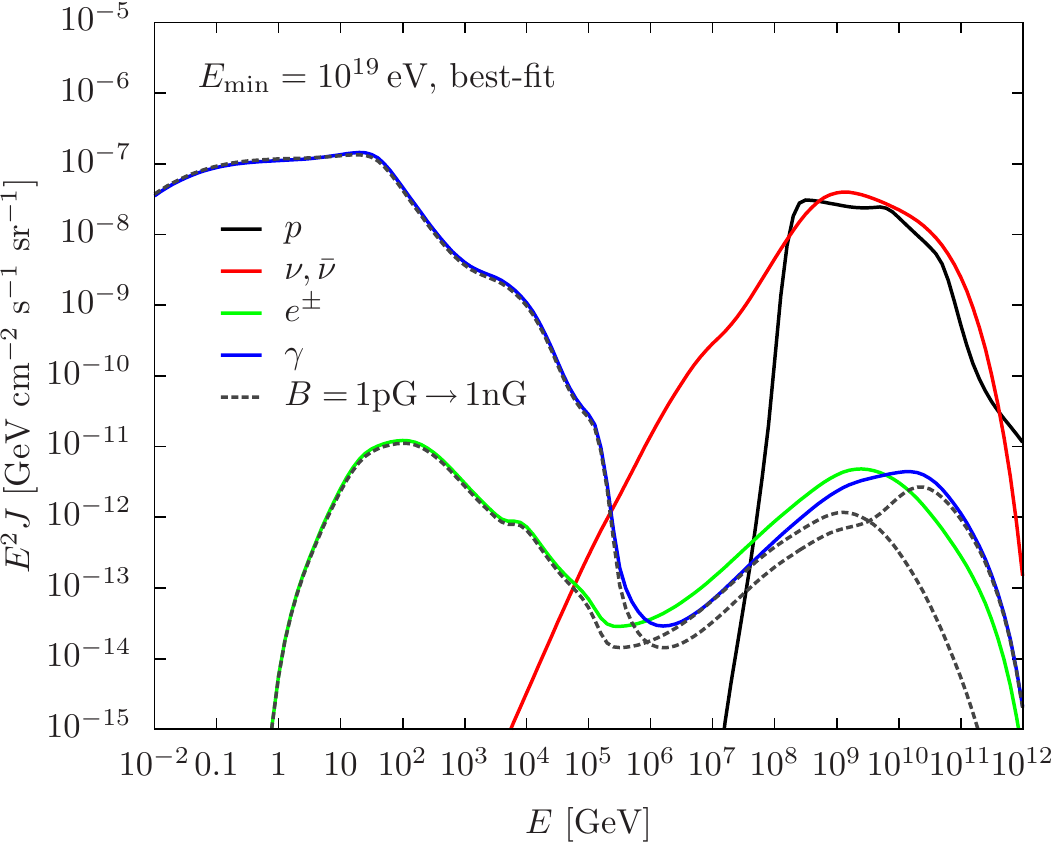}
\end{center}
\vspace{-0.5cm}
\caption[]{The effect of a large intergalactic magnetic field. The solid 
 lines show the spectra of the best-fit for crossover energies 
 $E_{\rm min}=10^{17.5}$~eV (left) and $10^{19}$~eV (right) using 
 $B_{\rm IG}=10^{-12}$~G. The dashed lines show the corresponding results 
 for a much larger field strength $B_{\rm IG}=10^{-9}$~G. The $\gamma$-ray 
 flux in the GeV-TeV region relevant for the Fermi-LAT spectrum is 
 practically unaffected.}\label{fig:magplot}
\end{figure}


\begin{thebibliography}{99}

\bibitem{Penzias:1965wn}
  A.~A.~Penzias and R.~W.~Wilson,
  Astrophys.\ J.\  {\bf 142}, 419 (1965).

\bibitem{Greisen:1966jv}
  K.~Greisen,
  Phys.\ Rev.\ Lett.\  {\bf 16}, 748 (1966).

\bibitem{Zatsepin:1966jv}
  G.~T.~Zatsepin and V.~A.~Kuzmin,
  JETP Lett.\  {\bf 4}, 78 (1966)
  [Pisma Zh.\ Eksp.\ Teor.\ Fiz.\  {\bf 4}, 114 (1966)].

\bibitem{Beresinsky:1969qj}
  V.~S.~Beresinsky and G.~T.~Zatsepin,
  Phys.\ Lett.\  B {\bf 28}, 423 (1969).

\bibitem{Abbasi:2007sv}
  R.~Abbasi {\it et al.}  [HiRes Collaboration],
  Phys.\ Rev.\ Lett.\  {\bf 100}, 101101 (2008)
  [arXiv:astro-ph/0703099].

\bibitem{Abraham:2008ru}
  J.~Abraham {\it et al.}  [Pierre Auger Collaboration],
  Phys.\ Rev.\ Lett.\  {\bf 101}, 061101 (2008)
  [arXiv:0806.4302 [astro-ph]].

\bibitem{Stecker:1978ah}
  F.~W.~Stecker,
  Astrophys.\ J.\  {\bf 228}, 919 (1979).

\bibitem{Yoshida:pt}
S.~Yoshida and M.~Teshima,
Prog.\ Theor.\ Phys.\  {\bf 89}, 833 (1993);
R.~J.~Protheroe and P.~A.~Johnson,
Astropart.\ Phys.\  {\bf 4}, 253 (1996) 
[arXiv:astro-ph/9506119];
R.~Engel, D.~Seckel and T.~Stanev,
Phys.\ Rev.\ D {\bf 64}, 093010 (2001)
[arXiv:astro-ph/0101216].

\bibitem{Hill:1985mk}
C.~T.~Hill and D.~N.~Schramm,
Phys.\ Rev.\ D {\bf 31}, 564 (1985).

\bibitem{Linsley:1963bk}
  J.~Linsley, {\it Proceedings of ICRC 1963, Jaipur, India}, pp.~77-99

\bibitem{Hill:1983mk}
  C.~T.~Hill and D.~N.~Schramm,
  Phys.\ Rev.\  D {\bf 31}, 564 (1985).

\bibitem{Wibig:2004ye}
  T.~Wibig and A.~W.~Wolfendale,
  J.\ Phys.\ G {\bf 31}, 255 (2005).
  [arXiv:astro-ph/0410624].
  
\bibitem{Berezinsky:2002nc}
  V.~Berezinsky, A.~Z.~Gazizov and S.~I.~Grigorieva,
  Phys.\ Rev.\  D {\bf 74}, 043005 (2006).
  [arXiv:hep-ph/0204357].
  
\bibitem{Fodor:2003ph} 
Z.~Fodor, S.~D.~Katz, A.~Ringwald and H.~Tu,
JCAP {\bf 0311}, 015 (2003)
[arXiv:hep-ph/0309171].

\bibitem{Hooper:2004jc}
D.~Hooper, A.~Taylor and S.~Sarkar,
Astropart.\ Phys.\  {\bf 23}, 11 (2005)
[arXiv:astro-ph/0407618];
M.~Ave, N.~Busca, A.~V.~Olinto, A.~A.~Watson and T.~Yamamoto,
Astropart.\ Phys.\  {\bf 23}, 19 (2005)
[arXiv:astro-ph/0409316];
 D.~Allard {\it et al.},
  JCAP {\bf 0609}, 005 (2006)
  [arXiv:astro-ph/0605327].

\bibitem{Anchordoqui:2007fi}
  L.~A.~Anchordoqui, H.~Goldberg, D.~Hooper, S.~Sarkar and A.~M.~Taylor,
  Phys.\ Rev.\  D {\bf 76}, 123008 (2007)
  [arXiv:0709.0734 [astro-ph]].
  We wish to stress that the essential results of this
  analysis are not altered by the new Auger
  data~\cite{Abraham:2010mj,Abraham:2010yv}.

\bibitem{Abraham:2010mj}
  J.~Abraham {\it et al.}  [Pierre Auger Collaboration],
  Phys.\ Lett.\  B {\bf 685}, 239 (2010)
  [arXiv:1002.1975 [astro-ph.HE]].

\bibitem{Abraham:2010yv}
  J.~Abraham {\it et al.}  [Pierre Auger Collaboration],
  Phys.\ Rev.\ Lett.\  {\bf 104}, 091101 (2010)
  [arXiv:1002.0699 [astro-ph.HE]]. 

\bibitem{Wibig:2008ji}
 T.~Wibig,
 arXiv:0810.5281 [hep-ph].

\bibitem{Ulrich:2009yq}
  R.~Ulrich, R.~Engel, S.~Muller, F.~Schussler and M.~Unger,
  Nucl.\ Phys.\ Proc.\ Suppl.\  {\bf 196}, 335 (2009)
  [arXiv:0906.3075 [astro-ph.HE]].

\bibitem{Abdo:2010nz}
  A.~A.~Abdo {\it et al.}  [Fermi-LAT Collaboration],
  Phys.\ Rev.\ Lett.\  {\bf 104}, 101101 (2010)
  [arXiv:1002.3603 [astro-ph.HE]].

\bibitem{Berezinsky:2010xa}
  V.~Berezinsky, A.~Gazizov, M.~Kachelriess and S.~Ostapchenko,
  arXiv:1003.1496 [astro-ph.HE].

\bibitem{Amsler:2008zzb}
  C.~Amsler {\it et al.}  [Particle Data Group],
  Phys.\ Lett.\  B {\bf 667}, 1 (2008).

\bibitem{Ahlers:2009rf}
  M.~Ahlers, L.~A.~Anchordoqui and S.~Sarkar,
  Phys.\ Rev.\  D {\bf 79}, 083009 (2009)
  [arXiv:0902.3993 [astro-ph.HE]].
  
\bibitem{Blumenthal:1970nn}
  G.~R.~Blumenthal,
  Phys.\ Rev.\  D {\bf 1}, 1596 (1970).

\bibitem{Mucke:1999yb}
  A.~M\"ucke, R.~Engel, J.~P.~Rachen, R.~J.~Protheroe and T.~Stanev,
  Comput.\ Phys.\ Commun.\  {\bf 124}, 290 (2000) 
  [arXiv:astro-ph/9903478].
  
\bibitem{Blumenthal:1970gc}
  G.~R.~Blumenthal and R.~J.~Gould,
  Rev.\ Mod.\ Phys.\  {\bf 42}, 237 (1970).

\bibitem{Lee:1996fp}
  S.~Lee,
  Phys.\ Rev.\  D {\bf 58}, 043004 (1998)
  [arXiv:astro-ph/9604098].

\bibitem{Demidov:2008az}
  S.~V.~Demidov and O.~E.~Kalashev,
  J.\ Exp.\ Theor.\ Phys.\  {\bf 108}, 764 (2009)
  [arXiv:0812.0859 [astro-ph]].

\bibitem{Kronberg:1993vk}
  P.~P.~Kronberg,
  Rept.\ Prog.\ Phys.\  {\bf 57}, 325 (1994).

\bibitem{Wdowczyk:1972}
  J.~Wdowczyk, W.~Tkaczyk and A.~W.~Wolfendale,
  J.~of Phys. A {\bf 5}, 1419-1432 (1972).

\bibitem{Dolag:2004kp}
  K.~Dolag, D.~Grasso, V.~Springel and I.~Tkachev,
  JCAP {\bf 0501}, 009 (2005)
  [arXiv:astro-ph/0410419].

\bibitem{Protheroe:1992dx}
  R.~J.~Protheroe and T.~Stanev,
  Mon. Not. R. Astron. Soc. {\bf 264}, 191 (1993).

\bibitem{Collaboration:2009uy}
  J.~Abraham {\it et al.}  [Pierre Auger Collaboration],
  Phys.\ Rev.\  D {\bf 79}, 102001 (2009)
  [arXiv:0903.3385 [astro-ph.HE]];
  Phys.\ Rev.\ Lett.\  {\bf 100}, 211101 (2008)
  [arXiv:0712.1909 [astro-ph]].

\bibitem{Achterberg:2007qp}
  A.~Achterberg {\it et al.}  [IceCube Collaboration],
  Phys.\ Rev.\  D {\bf 76}, 042008 (2007)
  [Erratum-ibid.\  D {\bf 77}, 089904 (2008)]
  [arXiv:0705.1315 [astro-ph]].

\bibitem{Ackermann:2007km}
  M.~Ackermann {\it et al.}  [IceCube Collaboration],
  Astrophys.\ J.\  {\bf 675}, 1014 (2008)
  [arXiv:0711.3022 [astro-ph]].
   
\bibitem{Aynutdinov:2005dq}
  V.~Aynutdinov {\it et al.}  [BAIKAL Collaboration],
  Astropart.\ Phys.\  {\bf 25}, 140 (2006)
  [arXiv:astro-ph/0508675].

\bibitem{Martens:2007ff}
  K.~Martens  [HiRes Collaboration],
  arXiv:0707.4417 [astro-ph].

\bibitem{Kravchenko:2006qc}
  I.~Kravchenko {\it et al.},
  Phys.\ Rev.\  D {\bf 73}, 082002 (2006)
  [arXiv:astro-ph/0601148].

\bibitem{Barwick:2005hn}
  S.~W.~Barwick {\it et al.}  [ANITA Collaboration],
  Phys.\ Rev.\ Lett.\  {\bf 96}, 171101 (2006)
  [arXiv:astro-ph/0512265].

\bibitem{Halzen:2006ic}
  F.~Halzen and D.~Hooper,
  Phys.\ Rev.\ Lett.\  {\bf 97}, 099901 (2006)
  [arXiv:astro-ph/0605103].

\bibitem{Achterberg:2006md}
  A.~Achterberg {\it et al.}  [IceCube Collaboration],
  Astropart.\ Phys.\  {\bf 26}, 155 (2006).
  
\bibitem{Ahrens:2003ix}
  J.~Ahrens {\it et al.}  [IceCube Collaboration],
  Astropart.\ Phys.\  {\bf 20}, 507 (2004)
  [arXiv:astro-ph/0305196].
  
\bibitem{Sreekumar:1997un}
  P.~Sreekumar {\it et al.}  [EGRET Collaboration],
  Astrophys.\ J.\  {\bf 494}, 523 (1998)
  [arXiv:astro-ph/9709257].

\bibitem{Strong:2004ry}
  A.~W.~Strong, I.~V.~Moskalenko and O.~Reimer,
  Astrophys.\ J.\  {\bf 613}, 956 (2004)
  [arXiv:astro-ph/0405441].

\bibitem{Anchordoqui:2009nf} 
  L.~A.~Anchordoqui and T.~Montaruli
  arXiv:0912.1035 [astro-ph.HE]. See Fig.~13 for an updated analysis
  of~\cite{Ahlers:2009rf}.

\bibitem{Ahlersetal}
  M.~Ahlers {\it et al.}, in preparation

\bibitem{Franceschini:2008tp}
 A.~Franceschini, G.~Rodighiero and M.~Vaccari,
 Astron.\ Astrophys.\  {\bf 487}, 837 (2008)
 [arXiv:0805.1841 [astro-ph]].

\bibitem{Aharonian:2005gh}
  F.~Aharonian {\it et al.}  [H.E.S.S. Collaboration],
  Nature {\bf 440}, 1018 (2006)
  [arXiv:astro-ph/0508073].

\bibitem{Aliu:2008ay}
  E.~Aliu {\it et al.} [MAGIC Collaboration],
  Science {\bf 320}, 1752 (2008)
  [arXiv:0807.2822 [astro-ph]].

\bibitem{LAT:2010kz}
  A.~A.~Abdo {\it et al.} [Fermi-LAT Collaboration],
  arXiv:1005.0996 [astro-ph.HE].

\bibitem{Bhattacharjee:1998qc}
  P.~Bhattacharjee and G.~Sigl,
  Phys.\ Rept.\  {\bf 327}, 109 (2000)
  [arXiv:astro-ph/9811011].

\bibitem{Clark} 
 T. A. Clark, L. W. Brown, and J. K. Alexander,
 Nature {\bf 228}, 847 (1970).

\bibitem{Protheroe:1996si}
R.~J.~Protheroe and P.~L.~Biermann,
Astropart.\ Phys.\  {\bf 6}, 45 (1996)
[Erratum-ibid.\  {\bf 7}, 181 (1996)],
[arXiv:astro-ph/9605119].

\bibitem{Sarkar:2003sp}
  S.~Sarkar,
  Acta Phys.\ Polon.\  B {\bf 35}, 351 (2004)
  [arXiv:hep-ph/0312223].

\bibitem{Schlickeiser:2002pg}
  R.~Schlickeiser,
  ``Cosmic ray astrophysics,''
{\it Berlin, Germany: Springer (2002) 519 p}

\bibitem{Kalashev:2007sn}
  O.~E.~Kalashev, D.~V.~Semikoz and G.~Sigl,
  Phys.\ Rev.\  D {\bf 79}, 063005 (2009)
  [arXiv:0704.2463 [astro-ph]].

\bibitem{Kelner:2008ke}
  S.~R.~Kelner and F.~A.~Aharonian,
  Phys.\ Rev.\  D {\bf 78}, 034013 (2008)
  [arXiv:0803.0688 [astro-ph]].
  
\end{thebibliography}
\end{document}